\documentclass[aps,prl,reprint,amsmath,amssymb,floatfix,citeautoscript,superscriptaddress]{revtex4-1}
\usepackage{amsmath}
\usepackage{graphicx}
\usepackage{amssymb}
\usepackage{amsthm}
\usepackage{bm}
\usepackage[breaklinks=true,colorlinks=true,linkcolor=blue,citecolor=blue,urlcolor=blue]{hyperref}

\usepackage{multirow}
\usepackage{times}
\usepackage{txfonts}
\usepackage{xfrac}

\usepackage{footnote}
\usepackage{afterpage}

\usepackage{color}

\begin{document}


\title{
First-Principles Approach for
Energy Level Alignment at Aqueous Semiconductor Interfaces}

\author{Neerav Kharche}
\email{nkharche@bnl.gov}
\affiliation{Department of Chemistry, Brookhaven National Laboratory, Upton, New York 11973-5000, USA}

\author{James T. Muckerman}
\email{muckerma@bnl.gov}
\affiliation{Department of Chemistry, Brookhaven National Laboratory, Upton, New York 11973-5000, USA}

\author{Mark S. Hybertsen}
\email{mhyberts@bnl.gov}
\affiliation{Center for Functional Nanomaterials, Brookhaven National Laboratory, Upton, New York 11973-5000, USA}

\begin{abstract}
A first-principles approach is demonstrated
to calculate the relationship between aqueous semiconductor interface
structure and energy level alignment.
The physical interface structure is sampled
using density functional theory based molecular dynamics,
yielding the interface electrostatic dipole.
The $GW$ approach is used to place the electronic band edge
energies of the semiconductor relative to the occupied $1b_1$ energy level in water.
Application to the specific cases of non-polar $(10\bar{1}0)$ facets 
of GaN and ZnO reveals a significant role for the structural motifs at the interface,
including the degree of interface water dissociation and the dynamical fluctuations
in the interface Zn-O and O-H bond orientations.
These effects contribute up to 0.5 eV.
\end{abstract}

\maketitle


The alignment of electronic energy levels at a heterointerface
between two materials
represents both a fundamental materials interface characteristic
and a crucial property that controls electronic device functionality
in such diverse areas as semiconductor electronics, batteries
and electrochemical cells.
In the case of semiconductor interfaces, the energy level
alignment is encapsulated in the valence band edge energy offset,
and a hierarchy of theoretical approaches
to calculate it have been established~\cite{Franciosi96}.
However, the corresponding energy level alignment at solid-electrolyte
interfaces poses a substantially more complex problem.
In particular, development of a constructive, first-principles theory 
for the alignment of semiconductor band-edge potentials 
to electrochemical potentials still presents fundamental challenges~\cite{Cheng12}.  

\begin{figure}[b]
\centering
\includegraphics[width=3.15in]{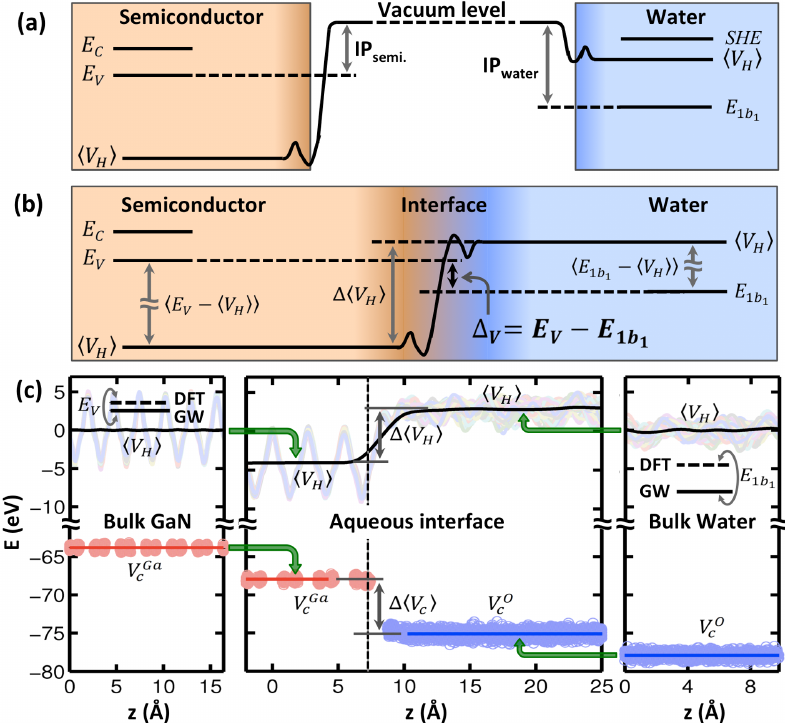}
\caption{
(a) Simplified model for energy level alignment based on
separate semiconductor and water energies with respect to vacuum.
(b) Schematic for energy level alignment including the
potential step due to the physical interface structure.
Valence band energy level offset $\Delta_V$ breaks
into three contributions in Eq.~\ref{eq:VBOffsetEq}
and shown here.
(c) Illustration of the separate calculation 
of each of those three 
terms for the
specific case of energy level alignment at 
the GaN$(10\bar{1}0)$-water interface.
Center panel: results of the
full interface simulation.
Top: planar, running and time (thermal) average 
of the electrostatic potential ($\langle V_H \rangle$, black line) 
superposed over 50 individual, 
planar-averaged 
snapshots.
Bottom: time and space averaged core potentials superposed over
individual values from the same snapshots.
Left panel: bulk GaN simulation
showing $E_V$ for GaN from DFT and $GW$
superposed on the averaged electrostatic potential
and the averaged core potentials. 
Right panel: bulk water showing 
$E_{1b_1}$ superposed on the same averaged potentials.
Final energy level alignment requires
shifting left and right panels in energy relative to the center panel
as illustrated by the arrows
to get the final alignment as in (b).
Either $V_H$ or equivalently the core potentials can be used for alignment.
}
\label{fig:scheme}
\end{figure}

As a significant example,
the relative alignment of the semiconductor band edge 
and the corresponding
redox level in the solvent for a target reaction 
determines thermodynamically whether
photoexcited carriers in the semiconductor 
can drive the reaction 
and with what range of overpotential.
This is of fundamental importance in the design of
electrochemical devices for solar energy harvesting~\cite{Nozik96, Kudo09}.
In particular, it is an unavoidable constraint in the
search for materials that can serve both as efficient
absorbers of the solar spectrum and to supply electrons
and holes with sufficient energy to drive relevant reactions,
e.g., the hydrogen evolution reaction or the water oxidation reaction.

In practical electrochemical measurements, the interface presents a complex
system: a doped semiconductor,
an aqueous interface of generally unknown atomic structure 
that may well participate in acid-base reactions
and the water with dissolved ions and a particular $\rm pH$.
Through control of the applied bias and the $\rm pH$,
conditions corresponding to both flat bands in the semiconductor
and zero zeta potential in the water can be achieved.
The latter refers to the condition of zero net interface charge
due to acid-base activity at the interface (adsorption of $\rm H^+$ and $\rm OH^-$)
~\cite{Cheng12}.
Under these conditions, the interface specific alignment of the
semiconductor band edges to the scale of redox levels in water
can be measured.  
In turn, this alignment determines semiconductor band bending
and water double-layer formation under general conditions of 
doping, ion concentration, bias and $\rm pH$.

In order to circumvent the complexity of these interfaces
and seek trends across semiconductors,
a simplified picture is appealing~\cite{Butler78}.
Imagine opening up the interface
so that one can characterize a reference semiconductor surface
and a water surface separately (Fig.~\ref{fig:scheme}a).
On the water side, the position of the highest occupied states
of bulk water ($1b_1$) are known from photoemission measurements~\cite{Winter04}.
The $\rm H^+/(\sfrac{1}{2})H_2$ redox level defines the standard hydrogen electrode (SHE)
and its absolute value relative to vacuum has been established~\cite{Cheng12,Trasatti86}.
On the semiconductor side, the ionization potential and electron affinity
fix the band edge positions relative to the vacuum.
Taken together, a model of the energy level alignment emerges.
\textit{What additional physical effects at the real interface
alter this simple picture and how large are they?}

In order to probe the impact of realistic semiconductor-water interface structure,
a constructive theory for the energy level alignment is required.
While distinct approaches have been explored~\cite{Cheng12,Cheng10,Wu11,ChengSelloni10},
at a key point in the analysis Kohn-Sham energy 
eigenvalues from Density Functional Theory (DFT)
are used to approximate electronic excitation energies.
This is well known to fail formally and practically, 
e.g., the well-known band gap problem~\cite{Jones89}.
The $GW$ approach in many-body perturbation theory
offers a well-founded theory for excitation energies~\cite{Hybertsen86,Godby87,Aulbur00}.
Recent applications of the $GW$ approach to liquid water
demonstrate substantial corrections for key electronic levels~\cite{Swartz13,Pham14}.
For trends in electrochemical energy level alignment,
based on the simplified model in Fig.~\ref{fig:scheme}a,
recent studies have incorporated corrections 
from the $GW$ approach~\cite{Toroker11,Li13,Stevanovic14}.
The broadest survey considered the calculated band edges,
the available photoemission data for the 
semiconductor surfaces, and electrochemical data~\cite{Stevanovic14}.
The authors infer that semiconductor-water interface structure
contributes about 0.5 eV to the alignment, 
albeit without explicit treatment of such structure.

Here we demonstrate a first-principles approach to
calculate the electronic excitation energy level alignment
at specific semiconductor-water interfaces 
($\Delta_V=E_V - E_{1b_1}$, Fig.~\ref{fig:scheme}b).
To do so, we must integrate dynamical sampling of the 
physical and electronic structure of both water
and the aqueous interface with the 
$GW$ approach for the excitation energies.

Our approach builds on the established methodology
used for semiconductor 
interfaces~\cite{VanDeWalle86,Zhang88,Hybertsen91,Franciosi96,Shaltaf08}.
DFT is utilized for the microscopic charge distribution at the interface, 
responsible for the interface-specific intrinsic dipole, while the
$GW$ approach is used for the calculation of the excitation energies
in the semiconductor and in the water.
Three separate calculations are required: 
one each for the bulk materials and a third for the interface properties.
The final energy level alignment combines
the results from these three calculations 
as illustrated in Fig.~\ref{fig:scheme}c,
where the suitably averaged electrostatic potential 
($V_H$) or core potential
is used for reference.

To represent the properties of the solid-liquid interface, 
\textit{ab initio} molecular dynamics (MD) are used to
simulate ambient conditions.
Electronic properties are determined by averaging a sample of
configurations to represent the thermal average and 
include the impact
of finite temperature renormalization of the energy levels,
albeit in a semiclassical approximation
~\cite{Lax52,Cardona05}.  
Our theory provides the semiconductor valence band alignment to the
centroid of the $1b_1$ band in liquid water
\begin{eqnarray}
\Delta_V = E_V - E_{1b_1} & = &
\langle E_V^{GW} - \langle V_H \rangle \rangle _{T,bulk} 
- \langle E_{1b_1}^{GW} 
- \langle V_H \rangle \rangle _{T,bulk} \nonumber \\
& + & \Delta \langle  \langle  V_H \rangle _{planar} \rangle _{T,interface} . 
\label{eq:VBOffsetEq}
\end{eqnarray}
This result links to the electrochemical scales in water
and the vacuum scale through the $E_{1b_1}$.

\begin{figure}[t]
\centering
\includegraphics{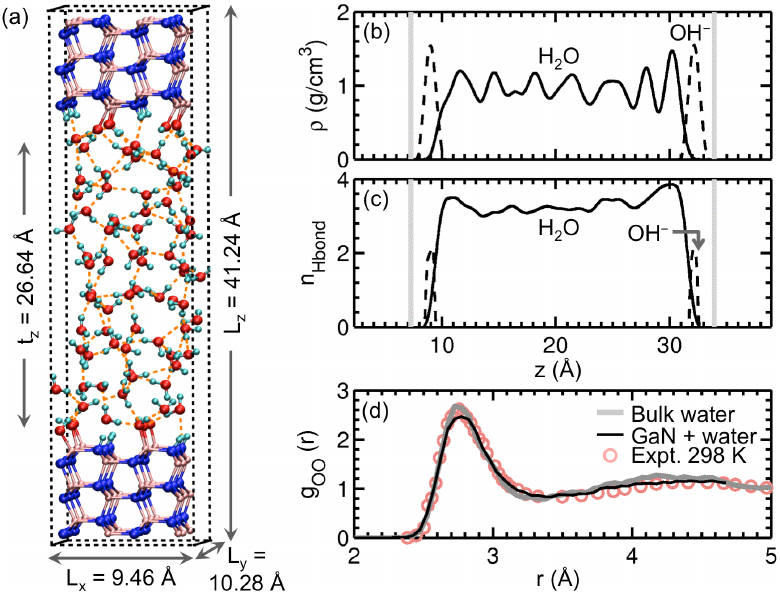}
\caption{GaN-water interface simulation. 
(a) Snapshot from the equilibrated portion of the MD simulation
illustrating the unit cell.
(b) Planar and time averaged density 
of water molecules (black solid line) and $\mathrm{OH^-}$ ions (black dashed line)
as a function of distance between the GaN interfaces (vertical gray lines). 
(c)  Average number of hydrogen bonds for water molecules (solid black line) and $\mathrm{OH^-}$ ions (black dashed line)
(d) Calculated O-O pair distribution function
from the bulk water simulation (gray line)
and the GaN interface simulation (black line) 
together with experiment~\cite{Soper08}.
}
\label{fig:interface}
\end{figure}

We demonstrate our approach for the 
specific cases of GaN and ZnO $(10\bar{1}0)$ 
interfaces with water, motivated by
the utility of GaN/ZnO mixed crystals for
photocatalysis with visible light~\cite{Maeda06}
and the recent observation that the $(10\bar{1}0)$
facet dominates the activity for GaN nanowires~\cite{Wang11a}.
We have recently analyzed the atomic-scale
structure at these aqueous semiconductor interfaces
using DFT-based MD simulations,
demonstrating the role of interface water dissociation~\cite{Kharche14}.
The same technical protocol is used here~\cite{Blochl94,Kresse96,Kresse99,Jensen08,Dion04,Klimes11,LandBorn},
including use of the projector augmented wave (PAW) method~\cite{Blochl94}
as implemented in VASP~\cite{Kresse96,Kresse99},
with the functional optB88-vdW~\cite{Dion04,Klimes11} 
that includes long-range van der Waals interactions.
See the Supplemental Material for details~\cite{SuppMat}.

A snapshot of the equilibrated structure
for the GaN case (Fig.~\ref{fig:interface}a)
shows that water near the interface with GaN spontaneously dissociates
resulting in a fully hydroxylated surface,
in agreement with prior work~\cite{Shen10,Wang11b,Kharche14}.
All surface N sites are protonated 
while all surface Ga sites are bonded to the
corresponding $\mathrm{OH^-}$ ions.
The average density of water, the characteristics
of the hydrogen bonds, the valence band density of states (DOS, not shown)
and the O-O pair distribution function
(excluding the near-surface regions) 
are close to those calculated for bulk water (Fig.~\ref{fig:interface}b-d)
and experiment~\cite{Soper08}.
For the ZnO interface,
the interface water layer is partially dissociated.
Half of the adsorbed water molecules dissociate for a $4\times2$ interface cell, 
in agreement with earlier studies~\cite{Dulub05,Tocci14,Kharche14}.
For a $3\times2$ cell, 
in-plane boundary conditions
result in fluctuations in the fraction of water dissociated
at the interface, including intervals where it is 67\%,
providing another structure for study.

To extract the average potential step between the semiconducting
region and the water region, the electrostatic potential profile
is calculated for 50 snapshots sampled from a 5 ps window. 
These are averaged laterally
to obtain the traces shown in Fig.~\ref{fig:scheme}c,
followed by a running average
and a time average. 
The resulting average electrostatic
potential is flat both in the center of the semiconducting region
and in the center of the water region;
interface effects are localized.
In practice, the PAW core potential,
illustrated in Fig.~\ref{fig:scheme}c,
provides an accurate, physically equivalent approach.
It is readily available for every MD snapshot
and averages are performed using 
all snapshots in the chosen time window.
Comparison of averages over a series of shorter time windows (1-2 ps each),
suggests an error bar of less than 0.1 eV due to sampling~\cite{SuppMat}.

\begin{figure}[t]
\centering
\includegraphics[width=2.8in]{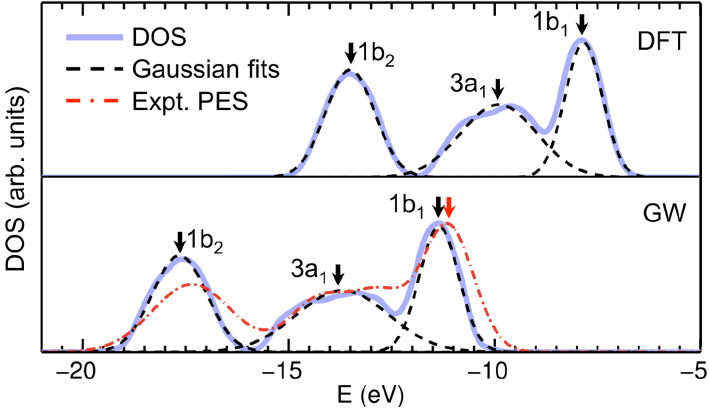}
\caption{
Valence band density of states (DOS) of bulk water,
relative to the vacuum level, calculated using
DFT (top) and $GW$ (bottom). 
Experimental PES spectra~\cite{Winter04} shown (bottom) 
with intensity scaled to match theory near the $1b_1$ peak.
}
\label{fig:dos}
\end{figure}

For the electronic excitation energies,
the $GW$ calculations are done with a full-frequency, 
spectrum-only self consistent approach, 
as implemented in VASP~\cite{Shishkin06,Shishkin07}.  
These specific choices in the method are based on previous results
for the energy band alignment at the $\mathrm{Si-SiO_2}$ interface
where this level of $GW$ self consistency gave
calculated band offsets within 0.3 eV of measured values ~\cite{Shaltaf08}.
Convergence with respect to the number of empty states
is achieved by a hyperbolic fit to a series of calculations
and extrapolation~\cite{Fredrich11,SuppMat}
For the case of water, electronic states 
of the 32-molecule cell
are analyzed for
a sample of 50 snapshots over a 5 ps period,
using $GW$ for a subset of 12 snapshots 
with the same energy gap distribution.

With this level of the theory, as expected~\cite{Shishkin07},
the calculated bulk band gaps of GaN (4.00 eV) and ZnO (3.93 eV)
are somewhat too large in comparison to experiment 
(3.44 and 3.3 eV, respectively)~\cite{Vurgaftman03,Sikant98},
all at room temperature~\cite{SuppMat}.
Also, as noted above, the optimized GaN and ZnO lattice constants 
are about 1\% smaller than the experimental values, 
which contributes a 0.2-0.3 eV increase to the band gap
through the deformation potentials~\cite{Vurgaftman03}.
For bulk water, the calculated average band gap
based on DFT is 4.35 eV while the present
$GW$ approach
gives 9.53 eV, slightly larger than the experimental 
value of $8.7\pm0.5$ eV~\cite{Bernas97}.
Our use of spectrum-only self consistency accounts
for the increase relative to the recent $G_0W_0$ result (8.1 eV~\cite{Pham14}).
Furthermore, the binding energies of the occupied $3a_1$ band
and $1b_2$ band relative to the $1b_1$ band (see Fig.~\ref{fig:dos})
are much more accurate in the present $GW$ calculations (2.34 eV and 6.30 eV)
than for the DFT energies (2.05 eV and 5.66 eV) in comparison
to photoemission experiments (2.34 eV and 6.21 eV)~\cite{Winter04}.

Next, we analyze the ingredients for the simple picture of Fig.~\ref{fig:scheme}a.
From the MD simulation of a water slab~\cite{SuppMat}
we find $\Delta \langle V_H \rangle = -3.32 \rm \ eV$ at the water-vacuum interface, 
in good agreement with earlier studies ~\cite{Leung10,Kathmann11,Pham14,Lucking14}.
Using this value, the calculated bulk DOS for water is aligned relative to the vacuum level
and compared with the experimental PES~\cite{Winter04} in Fig.~\ref{fig:dos}.
The binding energies are underestimated in DFT 
while the $GW$ corrected binding energies
are in very good agreement with the experimental data,
particularly the $1b_1$ level 
(peak centroid) at $-11.32$ eV, 
compared to the measured value of $-11.16$ eV~\cite{Winter04}.
For the clean GaN and ZnO $(10\bar{1}0)$ surfaces, including relaxation,
the valence band edge with respect to vacuum is
calculated to be $-6.98$ eV and $-8.08$ eV, respectively.
We make direct comparison to photoemission experiments 
for the closely related ZnO$(11\bar{2}0)$ surface,
for which we calculated $-8.14$ eV, in good agreement,
for an absolute energy, with
the measured value, $-7.82$ eV ~\cite{Swank67}.
See Ref.~\cite{Stevanovic14} for a broader survey.
From these calibration examples for both ZnO and water,
the highest occupied level is slightly too deep
relative to vacuum, an error that partially cancels in the
final theoretical results for the band alignment below.
Overall, this suggests that errors in the band alignment
at the aqueous interface will be similar to those found previously
for the $\mathrm{Si-SiO_2}$ interface
with the same level of $GW$ self consistency ~\cite{Shaltaf08}.

\begin{figure}[t]
\centering
\includegraphics[width=3.3in]{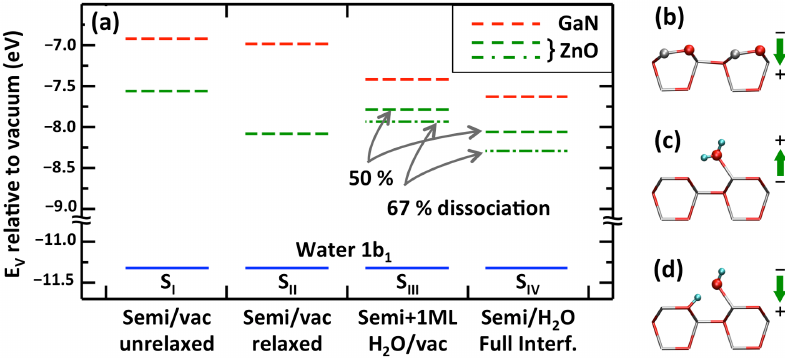}
\caption{
(a) Valence band edges, relative to vacuum, of GaN and ZnO for different surface
and interface configurations ($\rm S_I-S_{IV}$).
Solid horizontal lines depict the $1b_1$ level of water relative to vacuum.
For the full interface case, $\rm S_{IV}$, it is used
with the calculated offset $\Delta_V$ to place the
semiconductor valence band relative to vacuum.
Atomistic schematics with the qualitative induced surface
dipole shown (green arrows) due to:  
(b) relaxation of the clean ZnO surface, 
(c) molecular and (d) dissociative water adsorption, both on the ideal ZnO surface. 
}
\label{fig:alignment}
\end{figure}

The alignment of $E_V$ to the $1b_1$ level in water according to the simplified scheme is 
shown in Fig.~\ref{fig:alignment}a ($\rm S_{II}$).
The results for the ideal, unrelaxed semiconductor surfaces are also shown ($\rm S_{I}$).
Relaxation leads to rotation of the surface bond with outward (inward) displacement of the 
anion (cation) and an induced dipole
at the surface pointing inwards (Fig.~\ref{fig:alignment}b)
which lowers $E_V$ relative to vacuum.
This is a much larger effect for ZnO due to its more ionic bond.

Next we show the results from the full calculation,
including the structure of the semiconductor-water interface (Fig.~\ref{fig:scheme}b).
The calculated offsets ($\Delta_V = E_V-E_{1b_1}$)
are placed on the vacuum scale using the calculated value
for the $1b_1$ level from the water slab calculation (Fig.~\ref{fig:alignment}a, $\rm S_{IV}$).
For GaN, the change from the simple model is substantial,
with $E_V$ shifting to $-7.63$ eV.
For ZnO, the degree of interface water dissociation is quantitatively important,
with the final placement of $E_V$ being $-8.06$ eV and $-8.29$ eV for
50\% and 67\% dissociation, respectively.
In contrast to the GaN case, the simple model for ZnO is surprisingly close to the overall
result from the full calculation.
Analysis of the induced surface dipoles shows opposing effects
from the molecular and dissociative adsorption of water (Fig.~\ref{fig:alignment}c, d).
For GaN, lifting the surface reconstruction is dominated by dissociative adsorption of water
which lowers $E_V$.
For ZnO, the effect of lifting the reconstruction is larger while the
mixture of molecular and dissociative water adsorption compete,
with the latter being more significant in the 67\% dissociative case.

To obtain more insight into the role of interface structure motifs,
we show results in Fig.~\ref{fig:alignment}a ($\rm S_{III}$)
for each semiconductor with a single monolayer of adsorbed
water in vacuum, with the degree of dissociation found at the full interface.
The structures are fully relaxed, representative of a surface experiment.
The result for GaN, $-7.42$ eV, is rather close to the full calculation.
This suggests that the additional dipole induced by interaction of the hydroxylated
surface with liquid water is minimal.
On the other hand, the results for the ZnO case ($-7.79$ and $-7.94$ eV) agrees somewhat less
well with the full calculations.
Interestingly, key interface cation-O and O-H bond orientations fluctuate much
more in the ZnO case compared to the GaN case.
Correspondingly, the impact of the dipole contribution
from the liquid water interface to the hydroxylated surface is also larger,
another interface structure specific result.
However, our results do suggest that a well-chosen,
hydroxylated surface may be a better choice for application of the
simple picture in Fig.~\ref{fig:scheme}a for semiconductors
that actively promote water dissociation.

Electrochemical experiments for GaN~\cite{Kocha95,Beach03}
and for ZnO~\cite{Gomes82,Matsumoto89}
show acid-base activity at the semiconductor-water interface,
with a clear rise of the measured band edge positions with solution pH
(from 47 to 55 meV per pH unit, close to the ideal, 
Nernstian case of 59 meV).
Therefore, an additional measurement must determine the pH at which
the interface is neutral, assuring
the equivalent of flat-band conditions on the water side of the junction.
This has not been measured for GaN, but it is 
in the range of pH~= 8 to 10 for ZnO~\cite{Blok70,Kunze11}.
Taken together with the measured potential for the band edges~\cite{Gomes82,Matsumoto89},
the data are tightly clustered and relatively independent of facet,
covering a range of about 0.3 eV, centered on $E_V$ = $-7.3$ eV
relative to vacuum.
With reference to Fig.~\ref{fig:alignment}a,
the value of $E_V$ from experiment is at higher energy than that
predicted here for the ZnO $(10\bar{1}0)$ interface 
with a partially dissociated water layer. 
It is also
above the value from the simple picture of Fig.~\ref{fig:scheme}a,
(experimental value, $-7.82$ eV). 
Considering the impact on the dipole of different species at the interface
(Fig.~\ref{fig:alignment}), this suggests that under realistic
electrochemical conditions, another structural element with opposite
dipole to the net effect seen here must be involved.
Possibilities include alternative structures that result from etching
of the ZnO or the role of adsorption of other ions from solution,
both factors discussed in early literature~\cite{Gomes82,Blok70,Dewald60}.

In summary, we demonstrate the integrated use
of state-of-the-art techniques
for the first-principles treatment of energy level
alignment at aqueous semiconductor interfaces.
The initial, calibrated applications to 
GaN $(10\bar{1}0)$ and ZnO $(10\bar{1}0)$,
which exhibit different degrees of both water dissociation
and cation-O and O-H bond fluctuations at the interface, 
demonstrate the significant role of interface structure and dynamics.
In the future, this approach will support improved microscopic
understanding of chemical interactions and the impact of interface
structure on the fundamental energy alignments across semiconductor-water interfaces.\\

We thank P. B. Allen, M. Fernandez-Serra, D. Lu, and Y. Li for valuable discussions.
This work was carried out at Brookhaven National Laboratory 
under contract No. DE-AC02-98CH10886 with the U.S. Department of Energy, supported by its 
Office of Basic Energy Sciences (Computational Materials and Chemical Sciences Network program, 
Division of Chemical Sciences and Scientific User Facilities Division),
and utilized resources at the Center for Functional Nanomaterials, 
Brookhaven National Laboratory, and at the National Energy Research Scientific Computing Center, 
supported by the Office of Science of the U.S. Department of Energy 
under Contract No. DE-AC02-05CH11231.


%


\clearpage
\onecolumngrid
\section{Supplemental Material}
\setcounter{page}{1}

\subsection{Methods used and technical details for the calculations\hfill\hfill}
We have recently analyzed the atomic-scale
structure at aqueous GaN and ZnO $(10\bar{1}0)$ interfaces
using DFT-based MD simulations,
demonstrating the role of interface water dissociation~\cite{Kharche14}
and establishing the technical protocol used here.
DFT calculations are done 
using the projector augmented wave (PAW) method~\cite{Blochl94}
as implemented in VASP~\cite{Kresse96,Kresse99}.
The semicore $3d$ levels for Ga and Zn are explicitly treated
and the DFT+U method is
employed with $U_{Ga}$ = 3.9 eV and $U_{Zn}$ = 6.0 eV
for the $3d$ orbitals~\cite{Jensen08}.
The Born-Oppenheimer MD simulations are performed 
using a Nose-Hoover thermostat 
and a Verlet integrator with a time step of 0.5 fs.  
The functional optB88-vdW~\cite{Dion04,Klimes11}, 
including long-range van der Waals interactions, 
together with a slightly elevated temperature (T = 350 K), 
gives an accurate O-O pair distribution function 
and diffusivity. 
Converged results require a plane-wave cutoff energy of 600 eV.
All calculations are carried out at the optimized bulk lattice constants:
For GaN, $a = 3.154 ~\rm \AA$ and $c = 5.141 ~\rm \AA$ and 
for ZnO, $a = 3.208 ~\rm \AA$ and $c = 5.162 ~\rm \AA$.
These agree with the experimental values to within 1.3 \%~\cite{LandBorn}.
MD simulations are run for 10 to 20 ps
and analysis is performed for 5 ps thermalized windows.

The reference cell for bulk liquid water contains 32 molecules
(9.86 $\rm \AA^3$). 
The GaN-water interface is simulated using a repeated
supercell in which a 12 layer slab 
of $(10\bar{1}0)$ oriented wurtzite-structure
semiconductor with a $3\times2$ lateral cell
alternates with a water filled region containing 81 water molecules.
The statistical information in  (Fig.~2b-d) is obtained using 
the geometrical criteria described in our earlier work~\cite{Kharche14}.
Excluding 3 {\AA} thick surface layers, 
the O-O pair distribution function, hydrogen bond network, 
and valence band DOS are close to those calculated for bulk water.
The average density is slightly too large ($1.08 ~\rm g/cm^3$). 
The $1b_1$ band remained at the same energy relative to the core potential
due to cancellation in deformation potentials.
For the ZnO interface,
the interface water layer is partially dissociated.
Half of the adsorbed water molecules dissociate for a $4\times2$ interface cell
simulated with 66 water molecules.
For a $3\times2$ cell with 49 water molecules, 
in-plane boundary conditions
result in fluctuations in the fraction of water dissociated
at the interface, including intervals where it is 67\%,
providing another structure for study.
Finally, the water-vacuum interface is modeled using a supercell 
of the same size as that used for the GaN-water interface simulation.
In the supercell, an approximately 25 $\rm \AA$ thick water layer
containing 69 water molecules alternates 
with about 16 $\rm \AA$ thick vacuum region.. 

\setcounter{figure}{0}
\makeatletter 
\renewcommand{\thefigure}{S\@arabic\c@figure}
\makeatother

\begin{figure}[b]
\centering
\includegraphics[width=2.45in]{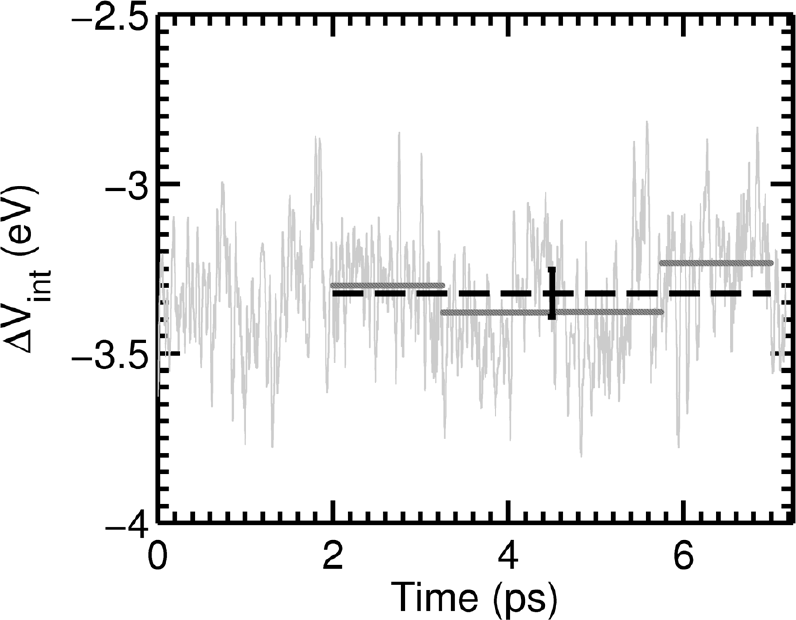}
\caption{
Electrostatic potential offset at the water-vacuum interface 
calculated using a water slab containing 69 water molecules 
for each snapshot from the MD simulation as a function of time. 
}
\label{fig:water_vac_offset}
\end{figure}

Calculations sampling the electronic structure from the
MD simulations naturally include the impact
of finite temperature renormalization of the energy levels,
albeit in a semiclassical approximation
~\cite{Lax52,Cardona05}.  
Although it is a relatively small effect in the semiconductors
at room temperature, for consistency with the treatment of liquid water,
we include them.
Specifically data from MD simulations for 12 layer, $3 \times 2$ bulk
supercells of GaN and ZnO
show a band gap renormalization 
of $-0.16$ and $-0.13$ eV respectively.

\begin{figure}[t]
\centering
\includegraphics[width=4.7in]{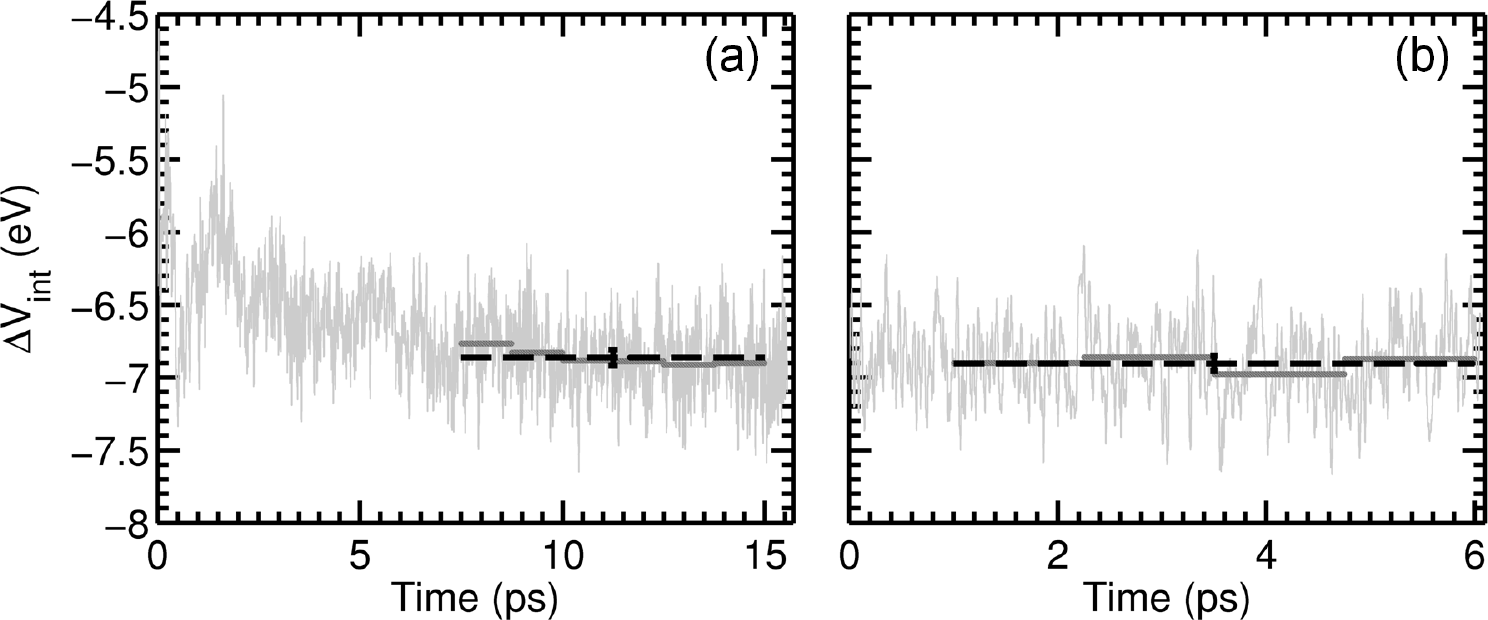}
\caption{
Electrostatic potential offset at the GaN-water interface calculated using a cell containing 
(a) $3\times2$ GaN slab and 48 water molecules and 
(b) $3\times2$ GaN slab and 81 water molecules.
Calculations for each snapshot from the MD simulation shown as a function of time.
}
\label{fig:GaN_water_offset}
\end{figure}

\subsection{Electrostatic potential offset as a function of the simulation time\hfill\hfill}
\begin{figure}[b]
\centering
\includegraphics[width=7in]{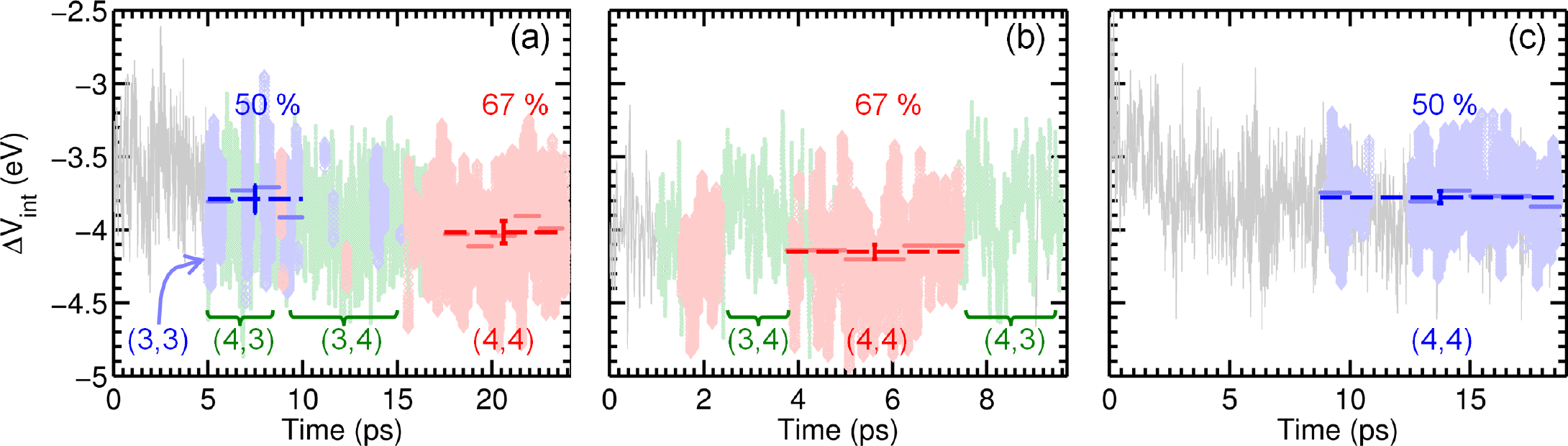}
\caption{
Electrostatic potential offset at the ZnO-water interface calculated using a cell containing 
(a) $3\times2$ ZnO slab and 49 water molecules, 
(b) $3\times2$ ZnO slab and 82 water molecules, 
and (c) $4\times2$ ZnO slab and 66 water molecules.
Calculations for each snapshot from the MD simulation shown as a function of time.
The traces are color coded according to the degree of water dissociation on the
top and bottom interfaces, noted in parentheses, e.g., (4, 3) indicates four dissociated
water molecules at the bottom interface per unit cell and three at the top one.
Blue corresponds to 50\% dissociation on top and bottom, red to 67\% dissociation top and bottom,
and green to structures where the degree of dissociation is different between top and bottom
resulting in electric fields within the water and semiconductor layers.
}
\label{fig:ZnO_water_offset}
\end{figure}

As described in the main text, we employ the PAW core potential as a reference 
to calculate the interface potential offset ($\Delta V_{int}$) as a function of the MD simulation time.
Specifically, the PAW core potentials are based on charge 
of radius (0.95, 1.03, 0.72)  {\AA} for (Zn, Ga, O).
Here we show the trajectories for all the aqueous interfaces studied in this work, 
Figs.~\ref{fig:water_vac_offset}-\ref{fig:ZnO_water_offset}, and provide further details
of the supercells studied and the impact of the thickness of the water region.
The potential offset for the clean semiconductor surfaces and the semiconductor surfaces with 1 ML of water adsorbed were calculated from a single, static structure with the relaxation specified in the text.  

The starting atomic configuration for the water-vacuum interface (Fig.~\ref{fig:water_vac_offset}) is obtained by 
removing the GaN slab and the adsorbed water molecules from a snapshot taken from 
the equilibrated portion of the MD trajectory of a cell containing $3\times2$ GaN slab and 81 water molecules.
This provides a well equilibrated water layer 
although with different surface boundary conditions.
The MD trajectory after 2 ps equilibration is used to calculate the electrostatic potential offsets 
by averaging all snapshots within 1.25 ps windows. 
In Fig.~\ref{fig:water_vac_offset} the values ($-3.30$, $-3.38$, $-3.38$, $-3.23$ eV) depicted by the horizontal solid grey lines 
average to $-3.32$ eV (black dashed line) with the standard deviation of 0.072 eV (vertical error bar).

For the data from the GaN-water interface in Fig.~\ref{fig:GaN_water_offset}, 
the same averaging scheme as that in Fig.~\ref{fig:water_vac_offset} is used.
The MD simulation is started with a smaller cell containing 48 water molecules. 
The starting atomic configuration for a larger cell containing 81 water molecules
is prepared by inserting a 33 water molecule slab at the ambient density 
in the middle of the water layer in an equilibrated snapshot of a smaller cell containing 48 water molecules.

In Fig.~\ref{fig:GaN_water_offset}a, 
the values ($-6.77$, $-6.83$, $-6.88$, $-6.89$, $-6.91$, $-6.90$ eV) 
depicted by the horizontal solid grey lines
average to -6.86 eV (black dashed line) with the standard deviation of 0.054 eV (vertical error bar).
In Fig.~\ref{fig:GaN_water_offset}b, the values ($-6.90$, $-6.86$, $-6.98$, $-6.87$ eV)  
depicted by the horizontal solid grey lines
average to $-6.90$ eV (black dashed line) with the standard deviation of 0.054 eV (vertical error bar).
Thus, the GaN-water electrostatic potential offset is well converged
with respect to the thickness of the water layer.

For the data from the ZnO-water interface in Fig.~\ref{fig:ZnO_water_offset}, 
the same averaging scheme as that in Fig.~\ref{fig:water_vac_offset} is used 
except only the snapshots corresponding to the symmetric 50 or 67\% water dissociation on both bottom and top interfaces are included.
Symmetric dissociation patterns are selected to avoid complications due to the
residual electric fields in the semiconductor slab and the bulk water region.
The MD simulation using $3\times2$ ($4\times2$) interface cell is started with a cell containing 49 (66) water molecules. 
The starting atomic configuration for a larger $3\times2$ cell containing 82 water molecules is obtained using 
the same procedure as described in Fig.~\ref{fig:GaN_water_offset}.

In Fig.~\ref{fig:ZnO_water_offset}a, the values ($-3.81$, $-3.73$, $-3.71$, $-3.92$ eV) 
for 50\% water dissociation depicted by the horizontal solid light blue lines
average to $-3.79$ eV (blue dashed line) 
with the standard deviation of  0.095 eV (vertical error bar) while 
the values ($-4.03$, $-4.11$, $-4.04$, $-3.91$, $-3.99$ eV) 
for 67\% water dissociation depicted by the horizontal solid light red lines
average to $-4.02$ eV (red dashed line) with the standard deviation of  0.073 eV (vertical error bar).
These data are used to report the comparison between 50\% and 67\% dissociated interfaces
in 
Fig.~4a 
of the main text.

In Fig.~\ref{fig:ZnO_water_offset}b, the values ($-4.14$, $-4.20$, $-4.11$ eV) 
for 67\% water dissociation depicted by the horizontal solid light red lines
average to $-4.15$ eV (red dashed line) with the standard deviation of  0.046 eV (vertical error bar).

In Fig.~\ref{fig:ZnO_water_offset}c, the values ($-3.75$, $-3.81$, $-3.73$, $-3.77$, $-3.77$, $-3.84$ eV) 
for 50\% water dissociation depicted by the horizontal solid light blue lines
average to $-3.78$ eV (blue dashed line) with the standard deviation of  0.040 eV (vertical error bar). 

Comparing the 67\% water dissociation in Fig.~\ref{fig:ZnO_water_offset}a and Fig.~\ref{fig:ZnO_water_offset}b, 
the ZnO-water electrostatic potential offset variation changes with respect to the thickness of the water layer by somewhat less than 0.15 eV.
Evidently the details of the transient stabilization of this interface structure depend on nearby water or proximity to the other interface.
Even though the interval of transient stabilization of 50\% water dissociation in Fig.~\ref{fig:ZnO_water_offset}a is smaller, 
the electrostatic potential offset agrees quite well with that found in Fig.~\ref{fig:ZnO_water_offset}c, 
for the $4\times2$ interface structure.
Without artificial constraints in the boundary conditions, the 50\% structure is likely more stable.

\begin{figure}[t]
\centering
\includegraphics[width=7in]{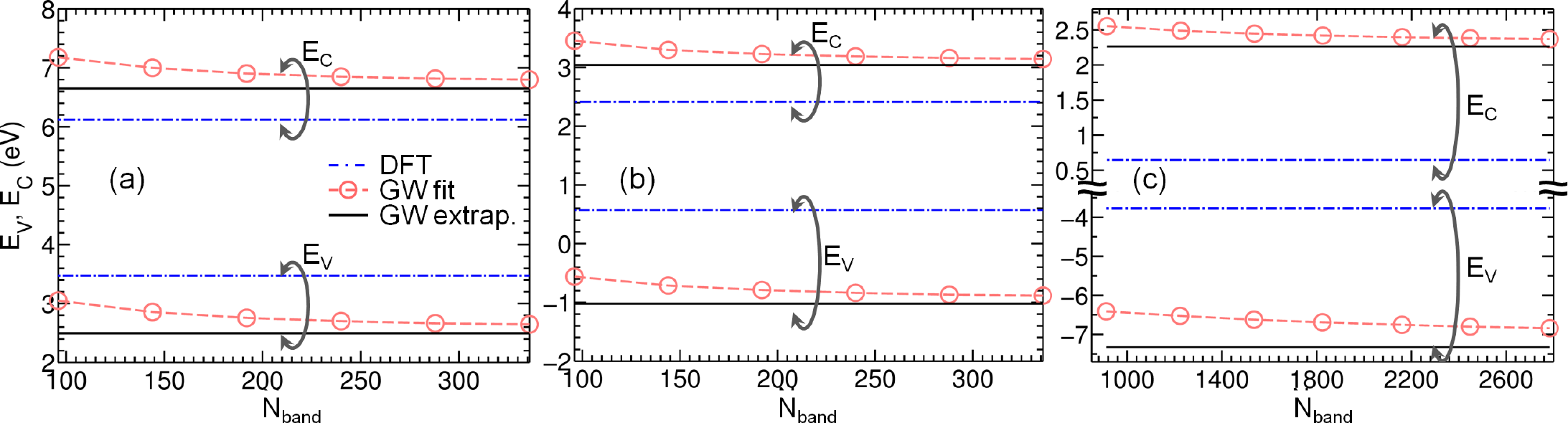}
\caption{
Convergence of the GW valence and conduction band edges with respect to the total number of bands for 
(a) GaN, (b) ZnO, and (c) one snapshot from MD simulation of water.
}
\label{fig:GW_nband_extrap}
\end{figure}

\subsection{Convergence of $GW$ results \hfill\hfill}
In the self-consistency, six cycles are sufficient for convergence to 0.05 eV.
An energy cutoff of 150 eV is used for the response function calculations.
We converged the $GW$ results relative to the number of empty states included 
using several calculations with ca. 5 to 20 times the number of occupied states 
together with a hyperbolic fit to extrapolate  to the infinite limit~\cite{Fredrich11}.
The hyperbolic function used is
\begin{equation}
f(N) = \frac{a}{N-N_0} + b
\label{eq:HyperbolicFit}
\end{equation}
where $N$ is the total number bands and $a$, $b$, and $N_0$ are fitting parameters.
The results are shown in Fig.~\ref{fig:GW_nband_extrap} 
for GaN, ZnO, and one snapshot from the MD simulation of the water cell containing 32 water molecules. 
To obtain the $GW$ density of states (DOS) of water shown in the bottom panel of Fig.~3 of the main text,
we used the same hyperbolic fitting method to extrapolate all valence band states
from 12 MD snapshots and averaged the results.
Note that the convergence of the valence and conduction band edges separately is 
considerably slower compared to the band gap.


\begin{thebibliography}{58}%
\makeatletter
\providecommand \@ifxundefined [1]{%
 \@ifx{#1\undefined}
}%
\providecommand \@ifnum [1]{%
 \ifnum #1\expandafter \@firstoftwo
 \else \expandafter \@secondoftwo
 \fi
}%
\providecommand \@ifx [1]{%
 \ifx #1\expandafter \@firstoftwo
 \else \expandafter \@secondoftwo
 \fi
}%
\providecommand \natexlab [1]{#1}%
\providecommand \enquote  [1]{``#1''}%
\providecommand \bibnamefont  [1]{#1}%
\providecommand \bibfnamefont [1]{#1}%
\providecommand \citenamefont [1]{#1}%
\providecommand \href@noop [0]{\@secondoftwo}%
\providecommand \href [0]{\begingroup \@sanitize@url \@href}%
\providecommand \@href[1]{\@@startlink{#1}\@@href}%
\providecommand \@@href[1]{\endgroup#1\@@endlink}%
\providecommand \@sanitize@url [0]{\catcode `\\12\catcode `\$12\catcode
  `\&12\catcode `\#12\catcode `\^12\catcode `\_12\catcode `\%12\relax}%
\providecommand \@@startlink[1]{}%
\providecommand \@@endlink[0]{}%
\providecommand \url  [0]{\begingroup\@sanitize@url \@url }%
\providecommand \@url [1]{\endgroup\@href {#1}{\urlprefix }}%
\providecommand \urlprefix  [0]{URL }%
\providecommand \Eprint [0]{\href }%
\providecommand \doibase [0]{http://dx.doi.org/}%
\providecommand \selectlanguage [0]{\@gobble}%
\providecommand \bibinfo  [0]{\@secondoftwo}%
\providecommand \bibfield  [0]{\@secondoftwo}%
\providecommand \translation [1]{[#1]}%
\providecommand \BibitemOpen [0]{}%
\providecommand \bibitemStop [0]{}%
\providecommand \bibitemNoStop [0]{.\EOS\space}%
\providecommand \EOS [0]{\spacefactor3000\relax}%
\providecommand \BibitemShut  [1]{\csname bibitem#1\endcsname}%
\let\auto@bib@innerbib\@empty
\bibitem [{\citenamefont {Franciosi}\ and\ \citenamefont {Van~de
  Walle}(1996)}]{Franciosi96}%
  \BibitemOpen
  \bibfield  {author} {\bibinfo {author} {\bibfnamefont {A.}~\bibnamefont
  {Franciosi}}\ and\ \bibinfo {author} {\bibfnamefont {C.~G.}\ \bibnamefont
  {Van~de Walle}},\ }\href@noop {} {\bibfield  {journal} {\bibinfo  {journal}
  {Surf. Sci. Rep.}\ }\textbf {\bibinfo {volume} {25}},\ \bibinfo {pages} {1}
  (\bibinfo {year} {1996})}\BibitemShut {NoStop}%
\bibitem [{\citenamefont {Cheng}\ and\ \citenamefont {Sprik}(2012)}]{Cheng12}%
  \BibitemOpen
  \bibfield  {author} {\bibinfo {author} {\bibfnamefont {J.}~\bibnamefont
  {Cheng}}\ and\ \bibinfo {author} {\bibfnamefont {M.}~\bibnamefont {Sprik}},\
  }\href@noop {} {\bibfield  {journal} {\bibinfo  {journal} {Phys. Chem. Chem.
  Phys.}\ }\textbf {\bibinfo {volume} {14}},\ \bibinfo {pages} {11245}
  (\bibinfo {year} {2012})}\BibitemShut {NoStop}%
\bibitem [{\citenamefont {Nozik}\ and\ \citenamefont
  {Memming}(1996)}]{Nozik96}%
  \BibitemOpen
  \bibfield  {author} {\bibinfo {author} {\bibfnamefont {A.~J.}\ \bibnamefont
  {Nozik}}\ and\ \bibinfo {author} {\bibfnamefont {R.}~\bibnamefont
  {Memming}},\ }\href@noop {} {\bibfield  {journal} {\bibinfo  {journal} {J.
  Phys. Chem.}\ }\textbf {\bibinfo {volume} {100}},\ \bibinfo {pages} {13061}
  (\bibinfo {year} {1996})}\BibitemShut {NoStop}%
\bibitem [{\citenamefont {Kudo}\ and\ \citenamefont {Miseki}(2009)}]{Kudo09}%
  \BibitemOpen
  \bibfield  {author} {\bibinfo {author} {\bibfnamefont {A.}~\bibnamefont
  {Kudo}}\ and\ \bibinfo {author} {\bibfnamefont {Y.}~\bibnamefont {Miseki}},\
  }\href@noop {} {\bibfield  {journal} {\bibinfo  {journal} {Chem. Soc. Rev.}\
  }\textbf {\bibinfo {volume} {38}},\ \bibinfo {pages} {253} (\bibinfo {year}
  {2009})}\BibitemShut {NoStop}%
\bibitem [{\citenamefont {Butler}\ and\ \citenamefont
  {Ginley}(1978)}]{Butler78}%
  \BibitemOpen
  \bibfield  {author} {\bibinfo {author} {\bibfnamefont {M.~A.}\ \bibnamefont
  {Butler}}\ and\ \bibinfo {author} {\bibfnamefont {D.~S.}\ \bibnamefont
  {Ginley}},\ }\href@noop {} {\bibfield  {journal} {\bibinfo  {journal} {J.
  Electrochem. Soc.}\ }\textbf {\bibinfo {volume} {125}},\ \bibinfo {pages}
  {228} (\bibinfo {year} {1978})}\BibitemShut {NoStop}%
\bibitem [{\citenamefont {Winter}\ \emph {et~al.}(2004)\citenamefont {Winter},
  \citenamefont {Weber}, \citenamefont {Widdra}, \citenamefont {Dittmar},
  \citenamefont {Faubel},\ and\ \citenamefont {Hertel}}]{Winter04}%
  \BibitemOpen
  \bibfield  {author} {\bibinfo {author} {\bibfnamefont {B.}~\bibnamefont
  {Winter}}, \bibinfo {author} {\bibfnamefont {R.}~\bibnamefont {Weber}},
  \bibinfo {author} {\bibfnamefont {W.}~\bibnamefont {Widdra}}, \bibinfo
  {author} {\bibfnamefont {M.}~\bibnamefont {Dittmar}}, \bibinfo {author}
  {\bibfnamefont {M.}~\bibnamefont {Faubel}}, \ and\ \bibinfo {author}
  {\bibfnamefont {I.~V.}\ \bibnamefont {Hertel}},\ }\href@noop {} {\bibfield
  {journal} {\bibinfo  {journal} {J. Phys. Chem. A}\ }\textbf {\bibinfo
  {volume} {108}},\ \bibinfo {pages} {2625} (\bibinfo {year}
  {2004})}\BibitemShut {NoStop}%
\bibitem [{\citenamefont {Trasatti}(1986)}]{Trasatti86}%
  \BibitemOpen
  \bibfield  {author} {\bibinfo {author} {\bibfnamefont {S.}~\bibnamefont
  {Trasatti}},\ }\href@noop {} {\bibfield  {journal} {\bibinfo  {journal} {Pure
  Appl. Chem.}\ }\textbf {\bibinfo {volume} {58}},\ \bibinfo {pages} {955}
  (\bibinfo {year} {1986})}\BibitemShut {NoStop}%
\bibitem [{\citenamefont {Cheng}\ and\ \citenamefont {Sprik}(2010)}]{Cheng10}%
  \BibitemOpen
  \bibfield  {author} {\bibinfo {author} {\bibfnamefont {J.}~\bibnamefont
  {Cheng}}\ and\ \bibinfo {author} {\bibfnamefont {M.}~\bibnamefont {Sprik}},\
  }\href@noop {} {\bibfield  {journal} {\bibinfo  {journal} {Phys. Rev. B}\
  }\textbf {\bibinfo {volume} {82}},\ \bibinfo {pages} {081406} (\bibinfo
  {year} {2010})}\BibitemShut {NoStop}%
\bibitem [{\citenamefont {Wu}\ \emph {et~al.}(2011)\citenamefont {Wu},
  \citenamefont {Chan},\ and\ \citenamefont {Ceder}}]{Wu11}%
  \BibitemOpen
  \bibfield  {author} {\bibinfo {author} {\bibfnamefont {Y.}~\bibnamefont
  {Wu}}, \bibinfo {author} {\bibfnamefont {M.~K.~Y.}\ \bibnamefont {Chan}}, \
  and\ \bibinfo {author} {\bibfnamefont {G.}~\bibnamefont {Ceder}},\
  }\href@noop {} {\bibfield  {journal} {\bibinfo  {journal} {Phys. Rev. B}\
  }\textbf {\bibinfo {volume} {83}},\ \bibinfo {pages} {235301} (\bibinfo
  {year} {2011})}\BibitemShut {NoStop}%
\bibitem [{\citenamefont {Cheng}\ and\ \citenamefont
  {Selloni}(2010)}]{ChengSelloni10}%
  \BibitemOpen
  \bibfield  {author} {\bibinfo {author} {\bibfnamefont {H.}~\bibnamefont
  {Cheng}}\ and\ \bibinfo {author} {\bibfnamefont {A.}~\bibnamefont
  {Selloni}},\ }\href@noop {} {\bibfield  {journal} {\bibinfo  {journal}
  {Langmuir}\ }\textbf {\bibinfo {volume} {26}},\ \bibinfo {pages} {11518}
  (\bibinfo {year} {2010})}\BibitemShut {NoStop}%
\bibitem [{\citenamefont {Jones}\ and\ \citenamefont
  {Gunnarsson}(1989)}]{Jones89}%
  \BibitemOpen
  \bibfield  {author} {\bibinfo {author} {\bibfnamefont {R.~O.}\ \bibnamefont
  {Jones}}\ and\ \bibinfo {author} {\bibfnamefont {O.}~\bibnamefont
  {Gunnarsson}},\ }\href@noop {} {\bibfield  {journal} {\bibinfo  {journal}
  {Rev. Mod. Phys.}\ }\textbf {\bibinfo {volume} {61}},\ \bibinfo {pages} {689}
  (\bibinfo {year} {1989})}\BibitemShut {NoStop}%
\bibitem [{\citenamefont {Hybertsen}\ and\ \citenamefont
  {Louie}(1986)}]{Hybertsen86}%
  \BibitemOpen
  \bibfield  {author} {\bibinfo {author} {\bibfnamefont {M.~S.}\ \bibnamefont
  {Hybertsen}}\ and\ \bibinfo {author} {\bibfnamefont {S.~G.}\ \bibnamefont
  {Louie}},\ }\href@noop {} {\bibfield  {journal} {\bibinfo  {journal} {Phys.
  Rev. B}\ }\textbf {\bibinfo {volume} {34}},\ \bibinfo {pages} {5390}
  (\bibinfo {year} {1986})}\BibitemShut {NoStop}%
\bibitem [{\citenamefont {Godby}\ \emph {et~al.}(1987)\citenamefont {Godby},
  \citenamefont {Schl{\"u}ter},\ and\ \citenamefont {Sham}}]{Godby87}%
  \BibitemOpen
  \bibfield  {author} {\bibinfo {author} {\bibfnamefont {R.~W.}\ \bibnamefont
  {Godby}}, \bibinfo {author} {\bibfnamefont {M.}~\bibnamefont {Schl{\"u}ter}},
  \ and\ \bibinfo {author} {\bibfnamefont {L.~J.}\ \bibnamefont {Sham}},\
  }\href@noop {} {\bibfield  {journal} {\bibinfo  {journal} {Phys. Rev. B}\
  }\textbf {\bibinfo {volume} {35}},\ \bibinfo {pages} {4170} (\bibinfo {year}
  {1987})}\BibitemShut {NoStop}%
\bibitem [{\citenamefont {Aulbur}\ \emph {et~al.}(2000)\citenamefont {Aulbur},
  \citenamefont {Jonsson},\ and\ \citenamefont {Wilkins}}]{Aulbur00}%
  \BibitemOpen
  \bibfield  {author} {\bibinfo {author} {\bibfnamefont {W.~G.}\ \bibnamefont
  {Aulbur}}, \bibinfo {author} {\bibfnamefont {L.}~\bibnamefont {Jonsson}}, \
  and\ \bibinfo {author} {\bibfnamefont {J.~W.}\ \bibnamefont {Wilkins}},\
  }\enquote {\bibinfo {title} {Quasiparticle calculations in solids},}\ in\
  \href@noop {} {\emph {\bibinfo {booktitle} {Solid State Physics, vol. 54}}},\
  \bibinfo {editor} {edited by\ \bibinfo {editor} {\bibfnamefont
  {H.}~\bibnamefont {Ehrenreich}}\ and\ \bibinfo {editor} {\bibfnamefont
  {F.}~\bibnamefont {Spaepen}}}\ (\bibinfo  {publisher} {Academic},\ \bibinfo
  {address} {New York},\ \bibinfo {year} {2000})\ pp.\ \bibinfo {pages}
  {1--218}\BibitemShut {NoStop}%
\bibitem [{\citenamefont {Swartz}\ and\ \citenamefont {Wu}(2013)}]{Swartz13}%
  \BibitemOpen
  \bibfield  {author} {\bibinfo {author} {\bibfnamefont {C.~W.}\ \bibnamefont
  {Swartz}}\ and\ \bibinfo {author} {\bibfnamefont {X.}~\bibnamefont {Wu}},\
  }\href@noop {} {\bibfield  {journal} {\bibinfo  {journal} {Phys. Rev. Lett.}\
  }\textbf {\bibinfo {volume} {111}},\ \bibinfo {pages} {087801} (\bibinfo
  {year} {2013})}\BibitemShut {NoStop}%
\bibitem [{\citenamefont {Pham}\ \emph {et~al.}(2014)\citenamefont {Pham},
  \citenamefont {Zhang}, \citenamefont {Schwegler},\ and\ \citenamefont
  {Galli}}]{Pham14}%
  \BibitemOpen
  \bibfield  {author} {\bibinfo {author} {\bibfnamefont {T.~A.}\ \bibnamefont
  {Pham}}, \bibinfo {author} {\bibfnamefont {C.}~\bibnamefont {Zhang}},
  \bibinfo {author} {\bibfnamefont {E.}~\bibnamefont {Schwegler}}, \ and\
  \bibinfo {author} {\bibfnamefont {G.}~\bibnamefont {Galli}},\ }\href@noop {}
  {\bibfield  {journal} {\bibinfo  {journal} {Phys. Rev. B}\ }\textbf {\bibinfo
  {volume} {89}},\ \bibinfo {pages} {060202} (\bibinfo {year}
  {2014})}\BibitemShut {NoStop}%
\bibitem [{\citenamefont {Toroker}\ \emph {et~al.}(2011)\citenamefont
  {Toroker}, \citenamefont {Kanan}, \citenamefont {Alidoust}, \citenamefont
  {Isseroff}, \citenamefont {Liao},\ and\ \citenamefont {Carter}}]{Toroker11}%
  \BibitemOpen
  \bibfield  {author} {\bibinfo {author} {\bibfnamefont {M.~C.}\ \bibnamefont
  {Toroker}}, \bibinfo {author} {\bibfnamefont {D.~K.}\ \bibnamefont {Kanan}},
  \bibinfo {author} {\bibfnamefont {N.}~\bibnamefont {Alidoust}}, \bibinfo
  {author} {\bibfnamefont {L.~Y.}\ \bibnamefont {Isseroff}}, \bibinfo {author}
  {\bibfnamefont {P.}~\bibnamefont {Liao}}, \ and\ \bibinfo {author}
  {\bibfnamefont {E.~A.}\ \bibnamefont {Carter}},\ }\href@noop {} {\bibfield
  {journal} {\bibinfo  {journal} {Phys. Chem. Chem. Phys.}\ }\textbf {\bibinfo
  {volume} {13}},\ \bibinfo {pages} {16644} (\bibinfo {year}
  {2011})}\BibitemShut {NoStop}%
\bibitem [{\citenamefont {Li}\ \emph {et~al.}(2013)\citenamefont {Li},
  \citenamefont {O{\rq}Leary}, \citenamefont {Lewis},\ and\ \citenamefont
  {Galli}}]{Li13}%
  \BibitemOpen
  \bibfield  {author} {\bibinfo {author} {\bibfnamefont {Y.}~\bibnamefont
  {Li}}, \bibinfo {author} {\bibfnamefont {L.~E.}\ \bibnamefont {O{\rq}Leary}},
  \bibinfo {author} {\bibfnamefont {N.~S.}\ \bibnamefont {Lewis}}, \ and\
  \bibinfo {author} {\bibfnamefont {G.}~\bibnamefont {Galli}},\ }\href@noop {}
  {\bibfield  {journal} {\bibinfo  {journal} {J. Phys. Chem. C}\ }\textbf
  {\bibinfo {volume} {117}},\ \bibinfo {pages} {5188} (\bibinfo {year}
  {2013})}\BibitemShut {NoStop}%
\bibitem [{\citenamefont {Stevanovic}\ \emph {et~al.}(2014)\citenamefont
  {Stevanovic}, \citenamefont {Lany}, \citenamefont {Ginley}, \citenamefont
  {Tumas},\ and\ \citenamefont {Zunger}}]{Stevanovic14}%
  \BibitemOpen
  \bibfield  {author} {\bibinfo {author} {\bibfnamefont {V.}~\bibnamefont
  {Stevanovic}}, \bibinfo {author} {\bibfnamefont {S.}~\bibnamefont {Lany}},
  \bibinfo {author} {\bibfnamefont {D.~S.}\ \bibnamefont {Ginley}}, \bibinfo
  {author} {\bibfnamefont {W.}~\bibnamefont {Tumas}}, \ and\ \bibinfo {author}
  {\bibfnamefont {A.}~\bibnamefont {Zunger}},\ }\href@noop {} {\bibfield
  {journal} {\bibinfo  {journal} {Phys. Chem. Chem. Phys.}\ }\textbf {\bibinfo
  {volume} {16}},\ \bibinfo {pages} {3706} (\bibinfo {year}
  {2014})}\BibitemShut {NoStop}%
\bibitem [{\citenamefont {Van~de Walle}\ and\ \citenamefont
  {Martin}(1986)}]{VanDeWalle86}%
  \BibitemOpen
  \bibfield  {author} {\bibinfo {author} {\bibfnamefont {C.~G.}\ \bibnamefont
  {Van~de Walle}}\ and\ \bibinfo {author} {\bibfnamefont {R.~M.}\ \bibnamefont
  {Martin}},\ }\href@noop {} {\bibfield  {journal} {\bibinfo  {journal} {Phys.
  Rev. B}\ }\textbf {\bibinfo {volume} {34}},\ \bibinfo {pages} {5621}
  (\bibinfo {year} {1986})}\BibitemShut {NoStop}%
\bibitem [{\citenamefont {Zhang}\ \emph {et~al.}(1988)\citenamefont {Zhang},
  \citenamefont {Tomanek}, \citenamefont {Louie}, \citenamefont {Cohen},\ and\
  \citenamefont {Hybertsen}}]{Zhang88}%
  \BibitemOpen
  \bibfield  {author} {\bibinfo {author} {\bibfnamefont {S.~B.}\ \bibnamefont
  {Zhang}}, \bibinfo {author} {\bibfnamefont {D.}~\bibnamefont {Tomanek}},
  \bibinfo {author} {\bibfnamefont {S.~G.}\ \bibnamefont {Louie}}, \bibinfo
  {author} {\bibfnamefont {M.~L.}\ \bibnamefont {Cohen}}, \ and\ \bibinfo
  {author} {\bibfnamefont {M.~S.}\ \bibnamefont {Hybertsen}},\ }\href@noop {}
  {\bibfield  {journal} {\bibinfo  {journal} {Solid State Comm.}\ }\textbf
  {\bibinfo {volume} {66}},\ \bibinfo {pages} {585} (\bibinfo {year}
  {1988})}\BibitemShut {NoStop}%
\bibitem [{\citenamefont {Hybertsen}(1991)}]{Hybertsen91}%
  \BibitemOpen
  \bibfield  {author} {\bibinfo {author} {\bibfnamefont {M.~S.}\ \bibnamefont
  {Hybertsen}},\ }\href@noop {} {\bibfield  {journal} {\bibinfo  {journal}
  {Appl. Phys. Lett.}\ }\textbf {\bibinfo {volume} {58}},\ \bibinfo {pages}
  {1759} (\bibinfo {year} {1991})}\BibitemShut {NoStop}%
\bibitem [{\citenamefont {Shaltaf}\ \emph {et~al.}(2008)\citenamefont
  {Shaltaf}, \citenamefont {Rignanese}, \citenamefont {Gonze}, \citenamefont
  {Giustino},\ and\ \citenamefont {Pasquarello}}]{Shaltaf08}%
  \BibitemOpen
  \bibfield  {author} {\bibinfo {author} {\bibfnamefont {R.}~\bibnamefont
  {Shaltaf}}, \bibinfo {author} {\bibfnamefont {G.~M.}\ \bibnamefont
  {Rignanese}}, \bibinfo {author} {\bibfnamefont {X.}~\bibnamefont {Gonze}},
  \bibinfo {author} {\bibfnamefont {F.}~\bibnamefont {Giustino}}, \ and\
  \bibinfo {author} {\bibfnamefont {A.}~\bibnamefont {Pasquarello}},\
  }\href@noop {} {\bibfield  {journal} {\bibinfo  {journal} {Phys. Rev. Lett.}\
  }\textbf {\bibinfo {volume} {100}},\ \bibinfo {pages} {186401} (\bibinfo
  {year} {2008})}\BibitemShut {NoStop}%
\bibitem [{\citenamefont {Lax}(1952)}]{Lax52}%
  \BibitemOpen
  \bibfield  {author} {\bibinfo {author} {\bibfnamefont {M.}~\bibnamefont
  {Lax}},\ }\href@noop {} {\bibfield  {journal} {\bibinfo  {journal} {J. Chem.
  Phys.}\ }\textbf {\bibinfo {volume} {20}},\ \bibinfo {pages} {1752} (\bibinfo
  {year} {1952})}\BibitemShut {NoStop}%
\bibitem [{\citenamefont {Cardona}\ and\ \citenamefont
  {Thewalt}(2005)}]{Cardona05}%
  \BibitemOpen
  \bibfield  {author} {\bibinfo {author} {\bibfnamefont {M.}~\bibnamefont
  {Cardona}}\ and\ \bibinfo {author} {\bibfnamefont {M.~L.~W.}\ \bibnamefont
  {Thewalt}},\ }\href@noop {} {\bibfield  {journal} {\bibinfo  {journal} {Rev.
  Mod. Phys.}\ }\textbf {\bibinfo {volume} {77}},\ \bibinfo {pages} {1173}
  (\bibinfo {year} {2005})}\BibitemShut {NoStop}%
\bibitem [{\citenamefont {Soper}\ and\ \citenamefont
  {Benmore}(2008)}]{Soper08}%
  \BibitemOpen
  \bibfield  {author} {\bibinfo {author} {\bibfnamefont {A.~K.}\ \bibnamefont
  {Soper}}\ and\ \bibinfo {author} {\bibfnamefont {C.~J.}\ \bibnamefont
  {Benmore}},\ }\href@noop {} {\bibfield  {journal} {\bibinfo  {journal} {Phys.
  Rev. Lett.}\ }\textbf {\bibinfo {volume} {101}},\ \bibinfo {pages} {065502}
  (\bibinfo {year} {2008})}\BibitemShut {NoStop}%
\bibitem [{\citenamefont {Maeda}\ \emph {et~al.}(2006)\citenamefont {Maeda},
  \citenamefont {Teramura}, \citenamefont {Lu}, \citenamefont {Takata},
  \citenamefont {Saito}, \citenamefont {Inoue},\ and\ \citenamefont
  {Domen}}]{Maeda06}%
  \BibitemOpen
  \bibfield  {author} {\bibinfo {author} {\bibfnamefont {K.}~\bibnamefont
  {Maeda}}, \bibinfo {author} {\bibfnamefont {K.}~\bibnamefont {Teramura}},
  \bibinfo {author} {\bibfnamefont {D.~L.}\ \bibnamefont {Lu}}, \bibinfo
  {author} {\bibfnamefont {T.}~\bibnamefont {Takata}}, \bibinfo {author}
  {\bibfnamefont {N.}~\bibnamefont {Saito}}, \bibinfo {author} {\bibfnamefont
  {Y.}~\bibnamefont {Inoue}}, \ and\ \bibinfo {author} {\bibfnamefont
  {K.}~\bibnamefont {Domen}},\ }\href@noop {} {\bibfield  {journal} {\bibinfo
  {journal} {Nature}\ }\textbf {\bibinfo {volume} {440}},\ \bibinfo {pages}
  {295} (\bibinfo {year} {2006})}\BibitemShut {NoStop}%
\bibitem [{\citenamefont {Wang}\ \emph
  {et~al.}(2011{\natexlab{a}})\citenamefont {Wang}, \citenamefont {Pierre},
  \citenamefont {Kibria}, \citenamefont {Cui}, \citenamefont {Han},
  \citenamefont {Bevan}, \citenamefont {Guo}, \citenamefont {Paradis},
  \citenamefont {Hakima},\ and\ \citenamefont {Mi}}]{Wang11a}%
  \BibitemOpen
  \bibfield  {author} {\bibinfo {author} {\bibfnamefont {D.~F.}\ \bibnamefont
  {Wang}}, \bibinfo {author} {\bibfnamefont {A.}~\bibnamefont {Pierre}},
  \bibinfo {author} {\bibfnamefont {M.~G.}\ \bibnamefont {Kibria}}, \bibinfo
  {author} {\bibfnamefont {K.}~\bibnamefont {Cui}}, \bibinfo {author}
  {\bibfnamefont {X.~G.}\ \bibnamefont {Han}}, \bibinfo {author} {\bibfnamefont
  {K.~H.}\ \bibnamefont {Bevan}}, \bibinfo {author} {\bibfnamefont
  {H.}~\bibnamefont {Guo}}, \bibinfo {author} {\bibfnamefont {S.}~\bibnamefont
  {Paradis}}, \bibinfo {author} {\bibfnamefont {A.~R.}\ \bibnamefont {Hakima}},
  \ and\ \bibinfo {author} {\bibfnamefont {Z.~T.}\ \bibnamefont {Mi}},\
  }\href@noop {} {\bibfield  {journal} {\bibinfo  {journal} {Nano Lett.}\
  }\textbf {\bibinfo {volume} {11}},\ \bibinfo {pages} {2353} (\bibinfo {year}
  {2011}{\natexlab{a}})}\BibitemShut {NoStop}%
\bibitem [{\citenamefont {Kharche}\ \emph {et~al.}(2014)\citenamefont
  {Kharche}, \citenamefont {Hybertsen},\ and\ \citenamefont
  {Muckerman}}]{Kharche14}%
  \BibitemOpen
  \bibfield  {author} {\bibinfo {author} {\bibfnamefont {N.}~\bibnamefont
  {Kharche}}, \bibinfo {author} {\bibfnamefont {M.~S.}\ \bibnamefont
  {Hybertsen}}, \ and\ \bibinfo {author} {\bibfnamefont {J.~T.}\ \bibnamefont
  {Muckerman}},\ }\href@noop {} {\bibfield  {journal} {\bibinfo  {journal}
  {Phys. Chem. Chem. Phys.}\ }\textbf {\bibinfo {volume} {16}},\ \bibinfo
  {pages} {12057} (\bibinfo {year} {2014})}\BibitemShut {NoStop}%
\bibitem [{\citenamefont {Bl{\"o}chl}(1994)}]{Blochl94}%
  \BibitemOpen
  \bibfield  {author} {\bibinfo {author} {\bibfnamefont {P.~E.}\ \bibnamefont
  {Bl{\"o}chl}},\ }\href@noop {} {\bibfield  {journal} {\bibinfo  {journal}
  {Phys. Rev. B}\ }\textbf {\bibinfo {volume} {50}},\ \bibinfo {pages} {17953}
  (\bibinfo {year} {1994})}\BibitemShut {NoStop}%
\bibitem [{\citenamefont {Kresse}\ and\ \citenamefont
  {Furthmuller}(1996)}]{Kresse96}%
  \BibitemOpen
  \bibfield  {author} {\bibinfo {author} {\bibfnamefont {G.}~\bibnamefont
  {Kresse}}\ and\ \bibinfo {author} {\bibfnamefont {J.}~\bibnamefont
  {Furthmuller}},\ }\href@noop {} {\bibfield  {journal} {\bibinfo  {journal}
  {Phys. Rev. B}\ }\textbf {\bibinfo {volume} {54}},\ \bibinfo {pages} {11169}
  (\bibinfo {year} {1996})}\BibitemShut {NoStop}%
\bibitem [{\citenamefont {Kresse}\ and\ \citenamefont
  {Joubert}(1999)}]{Kresse99}%
  \BibitemOpen
  \bibfield  {author} {\bibinfo {author} {\bibfnamefont {G.}~\bibnamefont
  {Kresse}}\ and\ \bibinfo {author} {\bibfnamefont {D.}~\bibnamefont
  {Joubert}},\ }\href@noop {} {\bibfield  {journal} {\bibinfo  {journal} {Phys.
  Rev. B}\ }\textbf {\bibinfo {volume} {59}},\ \bibinfo {pages} {1758}
  (\bibinfo {year} {1999})}\BibitemShut {NoStop}%
\bibitem [{\citenamefont {Jensen}\ \emph {et~al.}(2008)\citenamefont {Jensen},
  \citenamefont {Muckerman},\ and\ \citenamefont {Newton}}]{Jensen08}%
  \BibitemOpen
  \bibfield  {author} {\bibinfo {author} {\bibfnamefont {L.~L.}\ \bibnamefont
  {Jensen}}, \bibinfo {author} {\bibfnamefont {J.~T.}\ \bibnamefont
  {Muckerman}}, \ and\ \bibinfo {author} {\bibfnamefont {M.~D.}\ \bibnamefont
  {Newton}},\ }\href@noop {} {\bibfield  {journal} {\bibinfo  {journal} {J.
  Phys. Chem. C}\ }\textbf {\bibinfo {volume} {112}},\ \bibinfo {pages} {3439}
  (\bibinfo {year} {2008})}\BibitemShut {NoStop}%
\bibitem [{\citenamefont {Dion}\ \emph {et~al.}(2004)\citenamefont {Dion},
  \citenamefont {Rydberg}, \citenamefont {Schroder}, \citenamefont {Langreth},\
  and\ \citenamefont {Lundqvist}}]{Dion04}%
  \BibitemOpen
  \bibfield  {author} {\bibinfo {author} {\bibfnamefont {M.}~\bibnamefont
  {Dion}}, \bibinfo {author} {\bibfnamefont {H.}~\bibnamefont {Rydberg}},
  \bibinfo {author} {\bibfnamefont {E.}~\bibnamefont {Schroder}}, \bibinfo
  {author} {\bibfnamefont {D.~C.}\ \bibnamefont {Langreth}}, \ and\ \bibinfo
  {author} {\bibfnamefont {B.~I.}\ \bibnamefont {Lundqvist}},\ }\href@noop {}
  {\bibfield  {journal} {\bibinfo  {journal} {Phys. Rev. Lett.}\ }\textbf
  {\bibinfo {volume} {92}},\ \bibinfo {pages} {246401} (\bibinfo {year}
  {2004})}\BibitemShut {NoStop}%
\bibitem [{\citenamefont {Klimes}\ \emph {et~al.}(2011)\citenamefont {Klimes},
  \citenamefont {Bowler},\ and\ \citenamefont {Michaelides}}]{Klimes11}%
  \BibitemOpen
  \bibfield  {author} {\bibinfo {author} {\bibfnamefont {J.}~\bibnamefont
  {Klimes}}, \bibinfo {author} {\bibfnamefont {D.~R.}\ \bibnamefont {Bowler}},
  \ and\ \bibinfo {author} {\bibfnamefont {A.}~\bibnamefont {Michaelides}},\
  }\href@noop {} {\bibfield  {journal} {\bibinfo  {journal} {Phys. Rev. B}\
  }\textbf {\bibinfo {volume} {83}},\ \bibinfo {pages} {19513} (\bibinfo {year}
  {2011})}\BibitemShut {NoStop}%
\bibitem [{Lan()}]{LandBorn}%
  \BibitemOpen
  \href@noop {} {\bibinfo  {journal} {Springer Materials, Landolt-Bornstein
  Database}\ ,\ \bibinfo {pages}
  {http://www.springermaterials.com/docs/index.html}}\BibitemShut {NoStop}%
\bibitem [{Sup()}]{SuppMat}%
  \BibitemOpen
\bibfield  {journal} {  }\href@noop {} {}\bibinfo {note} {See Supplemental
  Material, which includes plots that illustrate the core potential offsets as
  a function of time and the extrapolation of the GW calculations versus number
  of empty states.}\BibitemShut {Stop}%
\bibitem [{\citenamefont {Shen}\ \emph {et~al.}(2010)\citenamefont {Shen},
  \citenamefont {Small}, \citenamefont {Wang}, \citenamefont {Allen},
  \citenamefont {Fernandez-Serra}, \citenamefont {Hybertsen},\ and\
  \citenamefont {Muckerman}}]{Shen10}%
  \BibitemOpen
  \bibfield  {author} {\bibinfo {author} {\bibfnamefont {X.~A.}\ \bibnamefont
  {Shen}}, \bibinfo {author} {\bibfnamefont {Y.~A.}\ \bibnamefont {Small}},
  \bibinfo {author} {\bibfnamefont {J.}~\bibnamefont {Wang}}, \bibinfo {author}
  {\bibfnamefont {P.~B.}\ \bibnamefont {Allen}}, \bibinfo {author}
  {\bibfnamefont {M.~V.}\ \bibnamefont {Fernandez-Serra}}, \bibinfo {author}
  {\bibfnamefont {M.~S.}\ \bibnamefont {Hybertsen}}, \ and\ \bibinfo {author}
  {\bibfnamefont {J.~T.}\ \bibnamefont {Muckerman}},\ }\href@noop {} {\bibfield
   {journal} {\bibinfo  {journal} {J. Phys. Chem. C}\ }\textbf {\bibinfo
  {volume} {114}},\ \bibinfo {pages} {13695} (\bibinfo {year}
  {2010})}\BibitemShut {NoStop}%
\bibitem [{\citenamefont {Wang}\ \emph
  {et~al.}(2011{\natexlab{b}})\citenamefont {Wang}, \citenamefont
  {Roman-Perez}, \citenamefont {Soler}, \citenamefont {Artacho},\ and\
  \citenamefont {Fernandez-Serra}}]{Wang11b}%
  \BibitemOpen
  \bibfield  {author} {\bibinfo {author} {\bibfnamefont {J.}~\bibnamefont
  {Wang}}, \bibinfo {author} {\bibfnamefont {G.}~\bibnamefont {Roman-Perez}},
  \bibinfo {author} {\bibfnamefont {J.~M.}\ \bibnamefont {Soler}}, \bibinfo
  {author} {\bibfnamefont {E.}~\bibnamefont {Artacho}}, \ and\ \bibinfo
  {author} {\bibfnamefont {M.~V.}\ \bibnamefont {Fernandez-Serra}},\
  }\href@noop {} {\bibfield  {journal} {\bibinfo  {journal} {J. Chem. Phys.}\
  }\textbf {\bibinfo {volume} {134}} (\bibinfo {year}
  {2011}{\natexlab{b}})}\BibitemShut {NoStop}%
\bibitem [{\citenamefont {Dulub}\ \emph {et~al.}(2005)\citenamefont {Dulub},
  \citenamefont {Meyer},\ and\ \citenamefont {Diebold}}]{Dulub05}%
  \BibitemOpen
  \bibfield  {author} {\bibinfo {author} {\bibfnamefont {O.}~\bibnamefont
  {Dulub}}, \bibinfo {author} {\bibfnamefont {B.}~\bibnamefont {Meyer}}, \ and\
  \bibinfo {author} {\bibfnamefont {U.}~\bibnamefont {Diebold}},\ }\href@noop
  {} {\bibfield  {journal} {\bibinfo  {journal} {Phys. Rev. Lett.}\ }\textbf
  {\bibinfo {volume} {95}},\ \bibinfo {pages} {136101} (\bibinfo {year}
  {2005})}\BibitemShut {NoStop}%
\bibitem [{\citenamefont {Tocci}\ and\ \citenamefont
  {Michaelides}(2014)}]{Tocci14}%
  \BibitemOpen
  \bibfield  {author} {\bibinfo {author} {\bibfnamefont {G.}~\bibnamefont
  {Tocci}}\ and\ \bibinfo {author} {\bibfnamefont {A.}~\bibnamefont
  {Michaelides}},\ }\href@noop {} {\bibfield  {journal} {\bibinfo  {journal}
  {J. Phys. Chem. Lett.}\ }\textbf {\bibinfo {volume} {5}},\ \bibinfo {pages}
  {474} (\bibinfo {year} {2014})}\BibitemShut {NoStop}%
\bibitem [{\citenamefont {Shishkin}\ and\ \citenamefont
  {Kresse}(2006)}]{Shishkin06}%
  \BibitemOpen
  \bibfield  {author} {\bibinfo {author} {\bibfnamefont {M.}~\bibnamefont
  {Shishkin}}\ and\ \bibinfo {author} {\bibfnamefont {G.}~\bibnamefont
  {Kresse}},\ }\href@noop {} {\bibfield  {journal} {\bibinfo  {journal} {Phys.
  Rev. B}\ }\textbf {\bibinfo {volume} {74}},\ \bibinfo {pages} {035101}
  (\bibinfo {year} {2006})}\BibitemShut {NoStop}%
\bibitem [{\citenamefont {Shishkin}\ and\ \citenamefont
  {Kresse}(2007)}]{Shishkin07}%
  \BibitemOpen
  \bibfield  {author} {\bibinfo {author} {\bibfnamefont {M.}~\bibnamefont
  {Shishkin}}\ and\ \bibinfo {author} {\bibfnamefont {G.}~\bibnamefont
  {Kresse}},\ }\href@noop {} {\bibfield  {journal} {\bibinfo  {journal} {Phys.
  Rev. B}\ }\textbf {\bibinfo {volume} {75}},\ \bibinfo {pages} {235102}
  (\bibinfo {year} {2007})}\BibitemShut {NoStop}%
\bibitem [{\citenamefont {Friedrich}\ \emph {et~al.}(2011)\citenamefont
  {Friedrich}, \citenamefont {Muller},\ and\ \citenamefont
  {Blugel}}]{Fredrich11}%
  \BibitemOpen
  \bibfield  {author} {\bibinfo {author} {\bibfnamefont {C.}~\bibnamefont
  {Friedrich}}, \bibinfo {author} {\bibfnamefont {M.~C.}\ \bibnamefont
  {Muller}}, \ and\ \bibinfo {author} {\bibfnamefont {S.}~\bibnamefont
  {Blugel}},\ }\href@noop {} {\bibfield  {journal} {\bibinfo  {journal} {Phys.
  Rev. B}\ }\textbf {\bibinfo {volume} {83}},\ \bibinfo {pages} {081101}
  (\bibinfo {year} {2011})}\BibitemShut {NoStop}%
\bibitem [{\citenamefont {Vurgaftman}\ and\ \citenamefont
  {Meyer}(2003)}]{Vurgaftman03}%
  \BibitemOpen
  \bibfield  {author} {\bibinfo {author} {\bibfnamefont {I.}~\bibnamefont
  {Vurgaftman}}\ and\ \bibinfo {author} {\bibfnamefont {J.~R.}\ \bibnamefont
  {Meyer}},\ }\href@noop {} {\bibfield  {journal} {\bibinfo  {journal} {J.
  Appl. Phys.}\ }\textbf {\bibinfo {volume} {94}},\ \bibinfo {pages} {3675}
  (\bibinfo {year} {2003})}\BibitemShut {NoStop}%
\bibitem [{\citenamefont {Srikant}\ and\ \citenamefont
  {Clarke}(1998)}]{Sikant98}%
  \BibitemOpen
  \bibfield  {author} {\bibinfo {author} {\bibfnamefont {V.}~\bibnamefont
  {Srikant}}\ and\ \bibinfo {author} {\bibfnamefont {D.~R.}\ \bibnamefont
  {Clarke}},\ }\href@noop {} {\bibfield  {journal} {\bibinfo  {journal} {J.
  Appl. Phys.}\ }\textbf {\bibinfo {volume} {83}},\ \bibinfo {pages} {5447}
  (\bibinfo {year} {1998})}\BibitemShut {NoStop}%
\bibitem [{\citenamefont {Bernas}\ \emph {et~al.}(1997)\citenamefont {Bernas},
  \citenamefont {Ferradini},\ and\ \citenamefont {Jay-Gerin}}]{Bernas97}%
  \BibitemOpen
  \bibfield  {author} {\bibinfo {author} {\bibfnamefont {A.}~\bibnamefont
  {Bernas}}, \bibinfo {author} {\bibfnamefont {C.}~\bibnamefont {Ferradini}}, \
  and\ \bibinfo {author} {\bibfnamefont {J.~P.}\ \bibnamefont {Jay-Gerin}},\
  }\href@noop {} {\bibfield  {journal} {\bibinfo  {journal} {Chem. Phys.}\
  }\textbf {\bibinfo {volume} {222}},\ \bibinfo {pages} {151} (\bibinfo {year}
  {1997})}\BibitemShut {NoStop}%
\bibitem [{\citenamefont {Leung}(2010)}]{Leung10}%
  \BibitemOpen
  \bibfield  {author} {\bibinfo {author} {\bibfnamefont {K.}~\bibnamefont
  {Leung}},\ }\href@noop {} {\bibfield  {journal} {\bibinfo  {journal} {J.
  Phys. Chem. Lett.}\ }\textbf {\bibinfo {volume} {1}},\ \bibinfo {pages} {496}
  (\bibinfo {year} {2010})}\BibitemShut {NoStop}%
\bibitem [{\citenamefont {Kathmann}\ \emph {et~al.}(2011)\citenamefont
  {Kathmann}, \citenamefont {Kuo}, \citenamefont {Mundy},\ and\ \citenamefont
  {Schenter}}]{Kathmann11}%
  \BibitemOpen
  \bibfield  {author} {\bibinfo {author} {\bibfnamefont {S.~M.}\ \bibnamefont
  {Kathmann}}, \bibinfo {author} {\bibfnamefont {I.-F.~W.}\ \bibnamefont
  {Kuo}}, \bibinfo {author} {\bibfnamefont {C.~J.}\ \bibnamefont {Mundy}}, \
  and\ \bibinfo {author} {\bibfnamefont {G.~K.}\ \bibnamefont {Schenter}},\
  }\href@noop {} {\bibfield  {journal} {\bibinfo  {journal} {J. Phys. Chem. B}\
  }\textbf {\bibinfo {volume} {115}},\ \bibinfo {pages} {4369} (\bibinfo {year}
  {2011})}\BibitemShut {NoStop}%
\bibitem [{\citenamefont {Lucking}\ \emph {et~al.}(2014)\citenamefont
  {Lucking}, \citenamefont {Sun}, \citenamefont {West},\ and\ \citenamefont
  {Zhang}}]{Lucking14}%
  \BibitemOpen
  \bibfield  {author} {\bibinfo {author} {\bibfnamefont {M.}~\bibnamefont
  {Lucking}}, \bibinfo {author} {\bibfnamefont {Y.-Y.}\ \bibnamefont {Sun}},
  \bibinfo {author} {\bibfnamefont {D.}~\bibnamefont {West}}, \ and\ \bibinfo
  {author} {\bibfnamefont {S.}~\bibnamefont {Zhang}},\ }\href@noop {}
  {\bibfield  {journal} {\bibinfo  {journal} {Chem. Sci.}\ }\textbf {\bibinfo
  {volume} {5}},\ \bibinfo {pages} {1216} (\bibinfo {year} {2014})}\BibitemShut
  {NoStop}%
\bibitem [{\citenamefont {Swank}(1967)}]{Swank67}%
  \BibitemOpen
  \bibfield  {author} {\bibinfo {author} {\bibfnamefont {R.~K.}\ \bibnamefont
  {Swank}},\ }\href@noop {} {\bibfield  {journal} {\bibinfo  {journal} {Phys.
  Rev.}\ }\textbf {\bibinfo {volume} {153}},\ \bibinfo {pages} {844} (\bibinfo
  {year} {1967})}\BibitemShut {NoStop}%
\bibitem [{\citenamefont {Kocha}\ \emph {et~al.}(1995)\citenamefont {Kocha},
  \citenamefont {Peterson}, \citenamefont {Arent}, \citenamefont {Redwing},
  \citenamefont {Tischler},\ and\ \citenamefont {Turner}}]{Kocha95}%
  \BibitemOpen
  \bibfield  {author} {\bibinfo {author} {\bibfnamefont {S.~S.}\ \bibnamefont
  {Kocha}}, \bibinfo {author} {\bibfnamefont {M.~W.}\ \bibnamefont {Peterson}},
  \bibinfo {author} {\bibfnamefont {D.~J.}\ \bibnamefont {Arent}}, \bibinfo
  {author} {\bibfnamefont {J.~M.}\ \bibnamefont {Redwing}}, \bibinfo {author}
  {\bibfnamefont {M.~A.}\ \bibnamefont {Tischler}}, \ and\ \bibinfo {author}
  {\bibfnamefont {J.~A.}\ \bibnamefont {Turner}},\ }\href@noop {} {\bibfield
  {journal} {\bibinfo  {journal} {J. Electrochem. Soc.}\ }\textbf {\bibinfo
  {volume} {142}},\ \bibinfo {pages} {L238} (\bibinfo {year}
  {1995})}\BibitemShut {NoStop}%
\bibitem [{\citenamefont {Beach}\ \emph {et~al.}(2003)\citenamefont {Beach},
  \citenamefont {Collins},\ and\ \citenamefont {Turner}}]{Beach03}%
  \BibitemOpen
  \bibfield  {author} {\bibinfo {author} {\bibfnamefont {J.~D.}\ \bibnamefont
  {Beach}}, \bibinfo {author} {\bibfnamefont {R.~T.}\ \bibnamefont {Collins}},
  \ and\ \bibinfo {author} {\bibfnamefont {J.~A.}\ \bibnamefont {Turner}},\
  }\href@noop {} {\bibfield  {journal} {\bibinfo  {journal} {J. Electrochem.
  Soc.}\ }\textbf {\bibinfo {volume} {150}},\ \bibinfo {pages} {A899} (\bibinfo
  {year} {2003})}\BibitemShut {NoStop}%
\bibitem [{\citenamefont {Gomes}\ and\ \citenamefont {Cardon}(1982)}]{Gomes82}%
  \BibitemOpen
  \bibfield  {author} {\bibinfo {author} {\bibfnamefont {W.~P.}\ \bibnamefont
  {Gomes}}\ and\ \bibinfo {author} {\bibfnamefont {F.}~\bibnamefont {Cardon}},\
  }\href@noop {} {\bibfield  {journal} {\bibinfo  {journal} {Progr. Surf.
  Sci.}\ }\textbf {\bibinfo {volume} {12}},\ \bibinfo {pages} {155} (\bibinfo
  {year} {1982})}\BibitemShut {NoStop}%
\bibitem [{\citenamefont {Matsumoto}\ \emph {et~al.}(1989)\citenamefont
  {Matsumoto}, \citenamefont {Yoshikawa},\ and\ \citenamefont
  {Sato}}]{Matsumoto89}%
  \BibitemOpen
  \bibfield  {author} {\bibinfo {author} {\bibfnamefont {Y.}~\bibnamefont
  {Matsumoto}}, \bibinfo {author} {\bibfnamefont {T.}~\bibnamefont
  {Yoshikawa}}, \ and\ \bibinfo {author} {\bibfnamefont {E.}~\bibnamefont
  {Sato}},\ }\href@noop {} {\bibfield  {journal} {\bibinfo  {journal} {J.
  Electrochem. Soc.}\ }\textbf {\bibinfo {volume} {136}},\ \bibinfo {pages}
  {1389} (\bibinfo {year} {1989})}\BibitemShut {NoStop}%
\bibitem [{\citenamefont {Blok}\ and\ \citenamefont {De~Bruyn}(1970)}]{Blok70}%
  \BibitemOpen
  \bibfield  {author} {\bibinfo {author} {\bibfnamefont {L.}~\bibnamefont
  {Blok}}\ and\ \bibinfo {author} {\bibfnamefont {P.~L.}\ \bibnamefont
  {De~Bruyn}},\ }\href@noop {} {\bibfield  {journal} {\bibinfo  {journal} {J.
  Colloid Interf. Sci.}\ }\textbf {\bibinfo {volume} {32}},\ \bibinfo {pages}
  {518} (\bibinfo {year} {1970})}\BibitemShut {NoStop}%
\bibitem [{\citenamefont {Kunze}\ \emph {et~al.}(2011)\citenamefont {Kunze},
  \citenamefont {Valtiner}, \citenamefont {Michels}, \citenamefont {Huber},\
  and\ \citenamefont {Grundmeier}}]{Kunze11}%
  \BibitemOpen
  \bibfield  {author} {\bibinfo {author} {\bibfnamefont {C.}~\bibnamefont
  {Kunze}}, \bibinfo {author} {\bibfnamefont {M.}~\bibnamefont {Valtiner}},
  \bibinfo {author} {\bibfnamefont {R.}~\bibnamefont {Michels}}, \bibinfo
  {author} {\bibfnamefont {K.}~\bibnamefont {Huber}}, \ and\ \bibinfo {author}
  {\bibfnamefont {G.}~\bibnamefont {Grundmeier}},\ }\href@noop {} {\bibfield
  {journal} {\bibinfo  {journal} {Phys. Chem. Chem. Phys.}\ }\textbf {\bibinfo
  {volume} {13}},\ \bibinfo {pages} {12959} (\bibinfo {year}
  {2011})}\BibitemShut {NoStop}%
\bibitem [{\citenamefont {Dewald}(1960)}]{Dewald60}%
  \BibitemOpen
  \bibfield  {author} {\bibinfo {author} {\bibfnamefont {J.~F.}\ \bibnamefont
  {Dewald}},\ }\href@noop {} {\bibfield  {journal} {\bibinfo  {journal} {J.
  Phys. Chem. Sol.}\ }\textbf {\bibinfo {volume} {14}},\ \bibinfo {pages} {155}
  (\bibinfo {year} {1960})}\BibitemShut {NoStop}%
\end{thebibliography}

\begin{thebibliography}{11}%
\makeatletter
\providecommand \@ifxundefined [1]{%
 \@ifx{#1\undefined}
}%
\providecommand \@ifnum [1]{%
 \ifnum #1\expandafter \@firstoftwo
 \else \expandafter \@secondoftwo
 \fi
}%
\providecommand \@ifx [1]{%
 \ifx #1\expandafter \@firstoftwo
 \else \expandafter \@secondoftwo
 \fi
}%
\providecommand \natexlab [1]{#1}%
\providecommand \enquote  [1]{``#1''}%
\providecommand \bibnamefont  [1]{#1}%
\providecommand \bibfnamefont [1]{#1}%
\providecommand \citenamefont [1]{#1}%
\providecommand \href@noop [0]{\@secondoftwo}%
\providecommand \href [0]{\begingroup \@sanitize@url \@href}%
\providecommand \@href[1]{\@@startlink{#1}\@@href}%
\providecommand \@@href[1]{\endgroup#1\@@endlink}%
\providecommand \@sanitize@url [0]{\catcode `\\12\catcode `\$12\catcode
  `\&12\catcode `\#12\catcode `\^12\catcode `\_12\catcode `\%12\relax}%
\providecommand \@@startlink[1]{}%
\providecommand \@@endlink[0]{}%
\providecommand \url  [0]{\begingroup\@sanitize@url \@url }%
\providecommand \@url [1]{\endgroup\@href {#1}{\urlprefix }}%
\providecommand \urlprefix  [0]{URL }%
\providecommand \Eprint [0]{\href }%
\providecommand \doibase [0]{http://dx.doi.org/}%
\providecommand \selectlanguage [0]{\@gobble}%
\providecommand \bibinfo  [0]{\@secondoftwo}%
\providecommand \bibfield  [0]{\@secondoftwo}%
\providecommand \translation [1]{[#1]}%
\providecommand \BibitemOpen [0]{}%
\providecommand \bibitemStop [0]{}%
\providecommand \bibitemNoStop [0]{.\EOS\space}%
\providecommand \EOS [0]{\spacefactor3000\relax}%
\providecommand \BibitemShut  [1]{\csname bibitem#1\endcsname}%
\let\auto@bib@innerbib\@empty
\bibitem [{\citenamefont {Kharche}\ \emph {et~al.}(2014)\citenamefont
  {Kharche}, \citenamefont {Hybertsen},\ and\ \citenamefont
  {Muckerman}}]{Kharche14}%
  \BibitemOpen
  \bibfield  {author} {\bibinfo {author} {\bibfnamefont {N.}~\bibnamefont
  {Kharche}}, \bibinfo {author} {\bibfnamefont {M.~S.}\ \bibnamefont
  {Hybertsen}}, \ and\ \bibinfo {author} {\bibfnamefont {J.~T.}\ \bibnamefont
  {Muckerman}},\ }\href@noop {} {\bibfield  {journal} {\bibinfo  {journal}
  {Phys. Chem. Chem. Phys.}\ }\textbf {\bibinfo {volume} {16}},\ \bibinfo
  {pages} {12057} (\bibinfo {year} {2014})}\BibitemShut {NoStop}%
\bibitem [{\citenamefont {Bl{\"o}chl}(1994)}]{Blochl94}%
  \BibitemOpen
  \bibfield  {author} {\bibinfo {author} {\bibfnamefont {P.~E.}\ \bibnamefont
  {Bl{\"o}chl}},\ }\href@noop {} {\bibfield  {journal} {\bibinfo  {journal}
  {Phys. Rev. B}\ }\textbf {\bibinfo {volume} {50}},\ \bibinfo {pages} {17953}
  (\bibinfo {year} {1994})}\BibitemShut {NoStop}%
\bibitem [{\citenamefont {Kresse}\ and\ \citenamefont
  {Furthmuller}(1996)}]{Kresse96}%
  \BibitemOpen
  \bibfield  {author} {\bibinfo {author} {\bibfnamefont {G.}~\bibnamefont
  {Kresse}}\ and\ \bibinfo {author} {\bibfnamefont {J.}~\bibnamefont
  {Furthmuller}},\ }\href@noop {} {\bibfield  {journal} {\bibinfo  {journal}
  {Phys. Rev. B}\ }\textbf {\bibinfo {volume} {54}},\ \bibinfo {pages} {11169}
  (\bibinfo {year} {1996})}\BibitemShut {NoStop}%
\bibitem [{\citenamefont {Kresse}\ and\ \citenamefont
  {Joubert}(1999)}]{Kresse99}%
  \BibitemOpen
  \bibfield  {author} {\bibinfo {author} {\bibfnamefont {G.}~\bibnamefont
  {Kresse}}\ and\ \bibinfo {author} {\bibfnamefont {D.}~\bibnamefont
  {Joubert}},\ }\href@noop {} {\bibfield  {journal} {\bibinfo  {journal} {Phys.
  Rev. B}\ }\textbf {\bibinfo {volume} {59}},\ \bibinfo {pages} {1758}
  (\bibinfo {year} {1999})}\BibitemShut {NoStop}%
\bibitem [{\citenamefont {Jensen}\ \emph {et~al.}(2008)\citenamefont {Jensen},
  \citenamefont {Muckerman},\ and\ \citenamefont {Newton}}]{Jensen08}%
  \BibitemOpen
  \bibfield  {author} {\bibinfo {author} {\bibfnamefont {L.~L.}\ \bibnamefont
  {Jensen}}, \bibinfo {author} {\bibfnamefont {J.~T.}\ \bibnamefont
  {Muckerman}}, \ and\ \bibinfo {author} {\bibfnamefont {M.~D.}\ \bibnamefont
  {Newton}},\ }\href@noop {} {\bibfield  {journal} {\bibinfo  {journal} {J.
  Phys. Chem. C}\ }\textbf {\bibinfo {volume} {112}},\ \bibinfo {pages} {3439}
  (\bibinfo {year} {2008})}\BibitemShut {NoStop}%
\bibitem [{\citenamefont {Dion}\ \emph {et~al.}(2004)\citenamefont {Dion},
  \citenamefont {Rydberg}, \citenamefont {Schroder}, \citenamefont {Langreth},\
  and\ \citenamefont {Lundqvist}}]{Dion04}%
  \BibitemOpen
  \bibfield  {author} {\bibinfo {author} {\bibfnamefont {M.}~\bibnamefont
  {Dion}}, \bibinfo {author} {\bibfnamefont {H.}~\bibnamefont {Rydberg}},
  \bibinfo {author} {\bibfnamefont {E.}~\bibnamefont {Schroder}}, \bibinfo
  {author} {\bibfnamefont {D.~C.}\ \bibnamefont {Langreth}}, \ and\ \bibinfo
  {author} {\bibfnamefont {B.~I.}\ \bibnamefont {Lundqvist}},\ }\href@noop {}
  {\bibfield  {journal} {\bibinfo  {journal} {Phys. Rev. Lett.}\ }\textbf
  {\bibinfo {volume} {92}},\ \bibinfo {pages} {246401} (\bibinfo {year}
  {2004})}\BibitemShut {NoStop}%
\bibitem [{\citenamefont {Klimes}\ \emph {et~al.}(2011)\citenamefont {Klimes},
  \citenamefont {Bowler},\ and\ \citenamefont {Michaelides}}]{Klimes11}%
  \BibitemOpen
  \bibfield  {author} {\bibinfo {author} {\bibfnamefont {J.}~\bibnamefont
  {Klimes}}, \bibinfo {author} {\bibfnamefont {D.~R.}\ \bibnamefont {Bowler}},
  \ and\ \bibinfo {author} {\bibfnamefont {A.}~\bibnamefont {Michaelides}},\
  }\href@noop {} {\bibfield  {journal} {\bibinfo  {journal} {Phys. Rev. B}\
  }\textbf {\bibinfo {volume} {83}},\ \bibinfo {pages} {19513} (\bibinfo {year}
  {2011})}\BibitemShut {NoStop}%
\bibitem [{Lan()}]{LandBorn}%
  \BibitemOpen
  \href@noop {} {\bibinfo  {journal} {Springer Materials, Landolt-Bornstein
  Database}\ ,\ \bibinfo {pages}
  {http://www.springermaterials.com/docs/index.html}}\BibitemShut {NoStop}%
\bibitem [{\citenamefont {Lax}(1952)}]{Lax52}%
  \BibitemOpen
\bibfield  {journal} {  }\bibfield  {author} {\bibinfo {author} {\bibfnamefont
  {M.}~\bibnamefont {Lax}},\ }\href@noop {} {\bibfield  {journal} {\bibinfo
  {journal} {J. Chem. Phys.}\ }\textbf {\bibinfo {volume} {20}},\ \bibinfo
  {pages} {1752} (\bibinfo {year} {1952})}\BibitemShut {NoStop}%
\bibitem [{\citenamefont {Cardona}\ and\ \citenamefont
  {Thewalt}(2005)}]{Cardona05}%
  \BibitemOpen
  \bibfield  {author} {\bibinfo {author} {\bibfnamefont {M.}~\bibnamefont
  {Cardona}}\ and\ \bibinfo {author} {\bibfnamefont {M.~L.~W.}\ \bibnamefont
  {Thewalt}},\ }\href@noop {} {\bibfield  {journal} {\bibinfo  {journal} {Rev.
  Mod. Phys.}\ }\textbf {\bibinfo {volume} {77}},\ \bibinfo {pages} {1173}
  (\bibinfo {year} {2005})}\BibitemShut {NoStop}%
\bibitem [{\citenamefont {Friedrich}\ \emph {et~al.}(2011)\citenamefont
  {Friedrich}, \citenamefont {Muller},\ and\ \citenamefont
  {Blugel}}]{Fredrich11}%
  \BibitemOpen
  \bibfield  {author} {\bibinfo {author} {\bibfnamefont {C.}~\bibnamefont
  {Friedrich}}, \bibinfo {author} {\bibfnamefont {M.~C.}\ \bibnamefont
  {Muller}}, \ and\ \bibinfo {author} {\bibfnamefont {S.}~\bibnamefont
  {Blugel}},\ }\href@noop {} {\bibfield  {journal} {\bibinfo  {journal} {Phys.
  Rev. B}\ }\textbf {\bibinfo {volume} {83}},\ \bibinfo {pages} {081101}
  (\bibinfo {year} {2011})}\BibitemShut {NoStop}%
\end{thebibliography}

%


\end{document}